\documentclass[aps,prb,twocolumn,nofootinbib,superscriptaddress,reprint,floatfix]{revtex4-2}
\usepackage{amsmath,amssymb,braket}
\usepackage{bbm}
\usepackage{ifpdf}
\usepackage{color}
\usepackage{xcolor}
\usepackage{graphicx,epsfig}
\usepackage{hyperref}
\usepackage{enumitem}
\usepackage{mathtools}
\usepackage{mathrsfs}

\hypersetup{
    colorlinks = True,
    linkcolor = {red},
    linkbordercolor = {red},
    citecolor = {blue},
    citebordercolor = {blue},
    urlcolor = {blue},
}

\usepackage[utf8]{inputenc}
\usepackage{tikz}
\usepackage{pgfplots} 
\usepackage{graphicx}
\usepackage{soul}

\usepackage[normalem]{ulem}

\pgfplotsset{compat=1.16} 
\begin{document}

   \title{Bose-Hubbard model in the canonical ensemble: a beyond mean-field approach}
    
   \author{Tista Banerjee}
   \email[]{banerjeemou09@gmail.com}
   \affiliation{School of Physical Sciences, Indian Association for the Cultivation of Science, Kolkata 700032, India}

     \begin{abstract}
        Ultracold atoms in optical lattices are versatile testbeds to study and manipulate equilibrium and out-of-equilibrium aspects of quantum many-body systems whose behavior can be described by Hubbard-type Hamiltonians. In this paper, we consider an ansatz wave-function which respects total particle-number conservation for such systems and goes beyond mean-field theory; this wave-function has the same complexity in the number of parameters as the mean-field Gutzwiller ansatz, and captures quantum correlations and entanglement via projection onto an effective low-energy manifold. This ansatz can be exploited to study quantum phases observed in a large class of systems realizable in such experimental platforms and is useful to study quantum dynamics. We show that the relaxation dynamics of various out-of-equilibrium initial states under sudden quench of Hamiltonian parameters can be studied with this ansatz wavefunction within the framework of time-dependent variational principle. We present a quantitative comparison with small-scale exact diagonalization results in the 1D Bose-Hubbard model with and without external trapping potentials. 
     \end{abstract}

    \maketitle

    \section{Introduction\label{sec:introduction}} Systems of ultra-cold atoms in an optical lattice offer a versatile platform to study and manipulate a host of equilibrium and out-of-equilibrium aspects of quantum many-body systems. When cold atoms are loaded onto an optical lattice, their behavior can be understood by considering only the lowest Bloch band and are described by Hubbard-type Hamiltonians. Depending on the atomic species under consideration, the form of the Hamiltonian can be an on-site or extended Bose-Hubbard model \cite{Zoller_1998_PhysRevLett.81.3108,Greiner2002}, Fermi-Hubbard model \cite{FHM_expt_PhysRevLett.94.080403} or variants of their mixtures \cite{Bose_Bose_mixtures_expt_PhysRevA.77.011603,Bose_Bose_mixtures_theory_PhysRevA.81.043620,Bose_fermi_mixtures_expt_PhysRevLett.96.180402} (see Ref.~\cite{Ueda_book} for a review).  \\
    
    One such model that describes strongly interacting lattice bosons in presence of repulsive onsite interaction is the paradigmatic Bose-Hubbard model. This model undergoes a Mott-insulator (MI) to superfluid (SF) transition when the depth of the optical lattice is tuned. This transition has been extensively studied both theoretically \cite{Fisher_1989_PhysRevB.40.546,GW_meanfield_K.Sheshadri_1993,Freericks_Monien_PhysRevB.53.2691,Kuhner_White_Monien_PhysRevB.61.12474,Carusotto_2003_YvanCastin,Sengupta_Dupuis_PhysRevA.71.033629,KS_twosp_PhysRevB.72.184507,Sansone_Prokofev_Svistunov_PhysRevB.75.134302,Pollet_Prokofev_Svistsunov_Troyer_PhysRevLett.103.140402,HRK_trivedi_PhysRevA.79.053631,KS_CT_PhysRevLett.106.095702,KS_CT_AD_PhysRevB.86.085140,Pai_PhysRevB.85.214524} and experimentally \cite{Esslinger_PhysRevLett.92.130403,Greiner_nature_2002,Ketterle_PhysRevLett.99.150604,Porto_PhysRevLett.100.120402}. \\
    
    A vast amount of theoretical studies on this model have relied on mean-field (MF) based methods to predict phases \cite{GW_meanfield_K.Sheshadri_1993,Fisher_1989_PhysRevB.40.546,Pai_PhysRevB.85.214524,Carusotto_2003_YvanCastin} and out-of-equilibrium dynamics \cite{YAN_2017_PhysRevA.95.053624,Natu_2010_PhysRevLett.106.125301,PhysRevLett.77.5315_CastinDum} (see Ref.~\cite{Kennet_nonequilibrioum_BHM_review_https://doi.org/10.1155/2013/393616} for a review). The preference for MF methods can be attributed to the simplicity and low computational complexity compared to alternative exact numerical approaches such as exact diagonalization (ED) \cite{ED_BHM_Zhang_2010} or quantum Monte Carlo (QMC) \cite{Pollet_2013_review}. Due to the large local Hilbert-space dimension for bosonic atoms or mixture of different atomic species, ED based approaches become quickly impractical due to memory requirements. While QMC methods are immune to this problem and have been extremely successful in studying equilibrium phases of Bose-Hubbard models \cite{Krauth_Trivedi_Ceperley_PhysRevLett.67.2307,Scaletta_Zimanyi_Kampf_PhysRevLett.74.2527,Prokofev1998,Sansone_Prokofev_Svistunov_PhysRevB.75.134302,Kuhner_White_Monien_PhysRevB.61.12474,Bloch_Dalibard_Zwerger_RevModPhys.80.885,Pollet_Prokofev_Svistsunov_Troyer_PhysRevLett.103.140402,Pollet_2013_review}, they are generally not well-suited for studying real-time quantum dynamics (although there have been some attempts \cite{QMC_real_time_2005_DOWLING2007549,QMC_real_time_2016_PhysRevLett.117.081602,Gull_PhysRevLett.115.266802} in the recent past).\\
    
    State-of-the art experiments often realize a canonical ensemble scenario where the number of bosonic atoms in the optical lattice do not change throughout the course of the experiment \cite{Bloch_nature_2010}. Most of the existing theoretical approaches \cite{GW_meanfield_K.Sheshadri_1993,GW_BHM_dynamics_Zakrzewski_PhysRevA.71.043601,Carusotto_2003_YvanCastin,YAN_2017_PhysRevA.95.053624,Natu_2010_PhysRevLett.106.125301,KS_CT_PhysRevLett.106.095702,KS_CT_AD_PhysRevB.86.085140,HRK_trivedi_PhysRevA.79.053631,Sengupta_Dupuis_PhysRevA.71.033629}, however, ignore this fact and essentially operate in a grand canonical ensemble scenario when probing their physical properties. \\
    
    A limited number of studies in the literature have focused on the role of this global $U(1)$ symmetry associated with particle-number conservation \cite{Number_projected_Gutz_Krauth_PhysRevB.45.3137,Rokhshar_generalised_GW_PhysRevB.44.10328,PhysRevA.57.3008_CastinDum,Zoller_atoic_matter_wave_revival_PhysRevA.83.043614,PhysRevA.78.053615_CastinSinatra,Pollet_canonical_ensemble_paper_PhysRevLett.96.180603} in determining the properties of these systems. In one such recent theoretical work \cite{Zoller_atoic_matter_wave_revival_PhysRevA.83.043614}, the collapse and revivals of atomic matter waves in an optical lattice was investigated using a particle-number-conserving Bose-Einstein condensate (BEC)-like ansatz. In this work it was shown that, although in the thermodynamic limit, this ansatz predicts the same quantitative results as the particle-number-non-conserving non-interacting BEC-like ansatz states described by the product of local coherent states over all sites, for small systems with finite numbers of particles, quantitative differences appear. Another extremely important effect is the recently observed dynamical restoration of the global symmetries \cite{cala1,cala2,cala7,cala8}. In this context, it is important to note that the dynamical restoration of the global $U(1)$ symmetry cannot be understood by particle-number-non-conserving approaches. In this case, it becomes imperative to work in the canonical ensemble by using particle number conserving approaches. \\

    Previously there have been attempts to incorporate $U(1)$ conservation within the Gutzwiller variational ansatz states \cite{Number_projected_Gutz_Krauth_PhysRevB.45.3137,Rokhshar_generalised_GW_PhysRevB.44.10328}, however, these are essentially MF in nature. On the other hand, the possibility of improving upon MF ansatz states by incorporating beyond MF correlations in a perturbative framework \cite{KS_CT_AD_PhysRevB.86.085140,KS_CT_PhysRevLett.106.095702,KS_twosp_PhysRevB.72.184507} have also been explored, but they are essentially $U(1)$ non-conserving. Our work aims to combine these two approaches to provide a framework for incorporating beyond MF correlations without giving up the $U(1)$ conservation. \\
     
    The goal of this paper is to formulate a beyond MF approach to study the ground state properties of ultra-cold bosonic systems, which manifestly preserves the global $U(1)$ symmetry and has relatively low computational complexity. We apply this beyond MF ansatz embedded in the canonical ensemble to the paradigmatic repulsive 1D Bose-Hubbard model (1DBHM) and show that it is able to correctly predict the ground-state (GS) properties of this model in regimes where the on-site interatomic interactions are much larger than the tunneling strength between neighboring sites. In order to benchmark the accuracy of this method, we present a comparison against ED in a broad range of scenarios. As we shall illustrate, this ansatz can accurately predict energy, condensate fraction, the presence of quasi long-range ordering and also notably the bipartite von Neumann entanglement entropy of the GS across a range of parameters. \\
    
    To emulate the ground-state properties accurately, we construct an ansatz wave-function by building correlations over a particle-number-projected MF Gutzwiller ansatz state by projecting it onto an effective low-energy manifold using appropriate canonical transformations. Such a construction is indeed possible in the strongly interacting regime. In particular, we employ perturbative Schrieffer-Wolff (SW) transformations to construct an effective low-energy Hamiltonian which is devoid of any hopping processes that cause an energy penalty of the order of on-site repulsive interaction strength in the lowest order of perturbation. \\
    
    Furthermore, we also present numerical evidence demonstrating that, apart from the GS properties, using such particle-number projected Gutzwiller ansatz states and their variants it is possible to reliably predict the out-of-equilibrium dynamics of these strongly correlated bosonic systems within the framework of time-dependent variational principle (TDVP). Importantly, since this approach operates entirely within the canonical ensemble, it becomes possible to study the dynamical restoration of the global $U(1)$ symmetry at the sub-system level and identify existence of a possible quantum Mpemba effect (QME). We present a quantitative comparison of TDVP and ED for the short-time dynamics of subsystem entanglement-asymmetry with respect to the global $U(1)$ symmetry as well as the sub-system evolution speed --- both of which quantities are useful for identification of anomalous relaxation processes --- starting from a class of initial states that explicitly break the global $U(1)$ symmetry. Our results indicate that altering the variational manifold from the well-known Gutzwiller-like state to a more correlated one opens up a path towards exploration of such anomalous relaxation mechanisms within the framework of TDVP. \\

    The key salient feature of methods explored in this paper are that while being moderately accurate, they are also computationally efficient. In order to probe ground-state properties, one only needs to optimize the energy expectation value of the system with respect to the ansatz wave-function. The energy expectation can be computed on-the-fly, and there are only a handful of parameters describing the ansatz wave-function. On the other hand, for probing the dynamical properties of the system, one needs to solve a set of coupled, non-linear, ordinary differential equations whose number is equal to the number of variational parameters required to specify the ansatz state. In our cases, the number of parameters required is the same as that of a Gutzwiller MF ansatz state. For a homogeneous system, this number scales with the local Hilbert space dimension, whereas for an inhomogeneous system, it scales as the product of local Hilbert space dimension and the size of the system. Due to the linear scaling of the number of variational parameters with the system size, this method remains feasible, even for moderate system sizes, when compared to alternate approaches like ED. The approaches discussed here are easily generalizable to higher dimensions and also to strongly correlated multi-species Bose-Hubbard models. \\

     The rest of this article is organized as follows. In Sec.~\ref{sec:model-method} we first introduce the 1D Bose-Hubbard model and outline our method. To benchmark this method, we consider various equilibrium and out-of-equilibrium scenarios. In Sec.~\ref{sec:equilibrium} we focus on the behavior of the energy, single particle Green's function and von Neumann entanglement entropy of a fixed number of atoms and with a homogeneous chemical potential. We also present results asserting the validity of the method in presence of an inhomogeneous chemical potential, in particular, a harmonic trapping potential. As we are working in the canonical ensemble, we briefly comment on the validity of the method in describing the Berezinskii-Kosterlitz-Thouless (BKT) transition at an integer filling. Finally, in Sec.~\ref{main:sec:dynamics} we present our results for the out-of-equilibrium dynamics. In particular, we compare the short-time dynamics of an initial BEC-like state using the time-dependent variational principle (TDVP) approach on the family of ansatz states introduced in Sec.~\ref{sec:model-method} against ED calculations. Here we consider BEC-like state in a deep optical lattice, instantaneous shift of a harmonic confining potential and quenches from a family of global $U(1)$ symmetry-broken initial states.


    \section{Model and Methods\label{sec:model-method}} For a one-dimensional lattice with periodic boundary conditions, the Bose-Hubbard model (1DBHM) Hamiltonian with on-site repulsive interaction reads 
    
    \begin{equation}
        \hat{H}= \sum_{\langle rr' \rangle } \left(-J\hat{b}^\dagger_r\hat{b}_{r'}+\text{H.c}\right) + \sum_{r=1}^L \left(-\mu_r\hat{n}_r+\frac{U_0}{2}\hat{n}_r\left(\hat{n}_r-1\right)\right)
    \label{eq:BHmodel-1D}
    \end{equation}

    Here $\hat{b}^\dagger_r$ ($\hat{b}_r$) are bosonic creation (annihilation) operators on the sites $r=1,2,3,...,L$ of a 1D lattice, $r'$ denotes the nearest neighbor of $r$, $J$ is the hopping strength, $\mu_r$ are chemical potentials which can be in principle site-dependent. The operator $\hat{n}_r \equiv \hat{b}_r^\dagger \hat{b}_r$ is the particle-number operator on site $r$ and $U_0(>0)$ is the strength of the on-site repulsive interaction. The Hamiltonian $\hat{H}$ has a global $U(1)$ symmetry corresponding to total particle-number-conservation, i.e. $[\hat{H},\hat{N}]=0$, where $\hat{N}$ is the total particle-number operator defined as $\hat{N}\equiv\sum_{r=1}^L \hat{n}_r$. The true ground-state (GS) of $\hat{H}$, must be an exact eigenstate of $\hat{N}$ --- a key aspect that is often overlooked in MF and MF-inspired studies of this model (Refs.~\cite{Number_projected_Gutz_Krauth_PhysRevB.45.3137,Rokhshar_generalised_GW_PhysRevB.44.10328} being important exceptions). One such well-known approach in the grand-canonical ensemble is the Gutzwiller MF (GWMF) method in which the variational ansatz $\ket{\Psi_{\text{GW}}(\boldsymbol{f})}$ takes the following form

    \begin{equation}
        \ket{\Psi_\text{GW}(\boldsymbol{f})} = \left[\bigotimes_{r=1}^L \left(\sum_{n_r=0}^{N_{\text{max}}}f_{r}^{n_r}\ket{n_r}\right)\right]
        \label{main:eq:gutz_ansatz}
    \end{equation}
    
    where, $\boldsymbol{f}=(f_1^0,f_1^1,...,f_L^{N_{\text{max}}})$ are parameters which are often referred to as Gutzwiller parameters/coefficients. The integer $N_{\text{max}}$ is a hard cut-off restricting the maximum number of allowed bosons per site, and this truncation of the local Hilbert space dimension is physically justified in regimes of strong interaction ($U_0\gg J$). The ansatz state $\ket{\Psi_{\text{GW}}(\boldsymbol{f})}$ is a product state of on-site superposition of different particle numbers, and is often used to study the GS properties \cite{GW_meanfield_K.Sheshadri_1993,KS_twosp_PhysRevB.72.184507,KS_CT_AD_PhysRevB.86.085140} and out-of-equilibrium dynamics \cite{GW_BHM_dynamics_Zakrzewski_PhysRevA.71.043601,KS_CT_AD_PhysRevB.86.085140,KS_CT_PhysRevLett.106.095702,Carusotto_2003_YvanCastin,YAN_2017_PhysRevA.95.053624,Natu_2010_PhysRevLett.106.125301} of the 1DBHM. However, the state $\ket{\Psi_{\text{GW}}(\boldsymbol{f})}$ is not an eigenstate of $\hat{N}$ for an arbitrary $\boldsymbol{f}$, and for reasons mentioned earlier, instead of the grand-canonical ensemble, our focus will be on the canonical ensemble. To ensure that we operate within the canonical ensemble, we perform a projection on to the fixed $N_b$-boson sector in the ansatz given in Eq.~\eqref{main:eq:gutz_ansatz} as

    \begin{equation}
        \ket{\tilde\Psi_{\text{GW}}(\boldsymbol{f})} = \mathcal{N}_{N_b} \hat{\mathcal{P}}_{N_b} \left[\bigotimes_{r=1}^L \left(\sum_{n_r=0}^{N_{\text{max}}}f_r^{n_r}\ket{n_r}\right)\right]
        \label{main:eq:fixed_nb_gutz_ansatz}
    \end{equation}

    Here $\hat{\mathcal{P}}_{N_b}$ is a projection operator corresponding to a desired number of total bosons $N_b$, and $\mathcal{N}_{N_b}$ is an overall factor which ensures the normalization $\langle\tilde\Psi_{\text{GW}}(\boldsymbol{f})|\tilde\Psi_{\text{GW}}(\boldsymbol{f})\rangle=1$. This ansatz has two features which appear due to the action of the projection operator $\hat{\mathcal{P}}_{N_b}$: (i) $\ket{\tilde\Psi_{\text{GW}}(\boldsymbol{f})}$ is now an exact eigenstate of the total particle number operator $\hat{N}$, which is a conserved charge of $\hat{H}$ of Eq.~\eqref{eq:BHmodel-1D}. (ii) The projection $\hat{\mathcal{P}}_{N_b}$ introduces non-trivial correlations, which entangles degrees of freedom at different sites of the lattice, unlike $\ket{\Psi_{\text{GW}}(\boldsymbol{f})}$ which is an unentangled state.

    One can now build additional correlations over this ansatz state $\ket{\tilde\Psi_{\text{GW}}(\boldsymbol{f})}$ in the same spirit as in Ref.~\cite{KS_CT_PhysRevLett.106.095702} by removing high-energy processes in \eqref{eq:BHmodel-1D} to some desired order in perturbation theory by appropriate canonical transformation and get a more accurate description of the GS wavefunction. To this end we note that if $\ket{\Psi}$ is an exact eigenstate of $\hat{H}$ then by performing a canonical transformation, the time-independent Schr\"odinger equation can be expressed as
    
    \begin{equation}
        \hat{H}_{\text{eff}} (e^{i\hat{S}}\ket{\Psi}) = E (e^{i\hat{S}}\ket{\Psi})
    \end{equation}

    Where $\hat{H}_\text{eff}=e^{i\hat{S}}\hat{H} e^{-i\hat{S}}$ is the effective Hamiltonian and $e^{i\hat{S}}\ket{\Psi}$ is the quantum state after the canonical transformation generated by $i\hat{S}$. This generator $i\hat{S}$ respects particle number conservation and removes hybridization between the eigenstates of $\hat{H}_0= \sum_{r=1}^L (-\mu_r\hat{n}_r+U_0\hat{n}_r\left(\hat{n}_r-1\right)/2)$ at $\mathcal{O}(J/U_0)$ in the strongly interacting regime. This can be achieved by considering the canonical transformation as a Schrieffer-Wolff (SW) transformation that eliminates the hopping processes which result in an $\mathcal{O}(U_0)$ change in the energy with respect to $\hat{H}_0$ and thus leads to an effective low-energy description of the system. We then make an assumption that this SW rotated ansatz $e^{i\hat{S}}\ket{\Psi}$ takes the form of a Gutzwiller ansatz state in the canonical ensemble (given by Eq.~\eqref{main:eq:fixed_nb_gutz_ansatz}). With such an assumption, the approximate wavefunction ($\ket{\tilde\Psi_{\text{SW}}(\boldsymbol{f})}$) of $\hat{H}$ is described as follows,

    \begin{equation}
        \ket{\tilde\Psi_{\text{SW}}(\boldsymbol{f})} = e^{-i\hat{S}} \mathcal{N}_{N_b} \hat{\mathcal{P}}_{N_b} \left[\bigotimes_{r=1}^L \left(\sum_{n_r=0}^{N_{\text{max}}}f_r^{n_r}\ket{n_r}\right)\right]
        \label{eq:fixed_nb_sw_gutz_ansatz}
    \end{equation}

    and the generator $i\hat{S}$, which satisfies the properties mentioned above can be shown to be of the following form (see Appendix-\ref{app:sec:SWrotHeff_derivation} for details)
    
  \begin{equation}
    \begin{split}
        i\hat{S}\ket{\{n\}} =& -J\sum_{r=1}^L \sum_{r'\in\text{nn}^{+}(r)} \frac{\hat{b}_r^\dagger \hat{b}_r'}{\Delta\epsilon_{rr'1}^{\alpha_{rr'}}} \delta(n_r-n_r'-\alpha_{rr'}+1) \ket{\{n\}} \\ &
        + \frac{\hat{b}_{r'}^\dagger \hat{b}_r}{\Delta\epsilon_{rr'2}^{\alpha_{rr'}}} \delta(n_r-n_r'+\alpha_{rr'}-1) \ket{\{n\}} + \mathcal{O}(\frac{J}{U_0})^2\label{eq:iS_defn}
    \end{split}
    \end{equation}    
    
    With this description, the approximate GS of the 1DBHM, in the strongly correlated regime can be obtained by computing the optimal set of parameters $\boldsymbol{f}$ in Eq.~\eqref{eq:fixed_nb_sw_gutz_ansatz} which minimizes the energy $E(\boldsymbol{f})= \langle \tilde\Psi_{\text{SW}}(\boldsymbol{f}) | \hat{H} | \tilde \Psi_{\text{SW}}(\boldsymbol{f}) \rangle = \langle \tilde\Psi_{\text{GW}}(\boldsymbol{f}) | \hat{H}_{\text{eff}} | \tilde \Psi_{\text{GW}}(\boldsymbol{f}) \rangle$. As we shall see, the approximate GS ($\ket{\tilde\Psi_{\text{SW}}(\boldsymbol{f})}$) obtained in this manner can better estimate the exact ground state properties of 1DBHM in the strongly correlated regime compared to ansatz Eq.~\eqref{main:eq:fixed_nb_gutz_ansatz}. This happens as the new state which incorporates correlations over particle-number projected Gutzwiller ansatz state (Eq.~\eqref{main:eq:fixed_nb_gutz_ansatz}), arises from an extra projection on the an effective low-energy manifold via the aforementioned SW transformation. The upshot of all of this is that we have computed an effective Hamiltonian up to $\mathcal{O}(J/U_0)^2$ which, via hopping processes, induces correlation ranging over $3$ lattice sites from any given site, in the approximate wavefunction (see Appendix-\ref{app:sec:SWrotHeff_derivation} for details). As we shall illustrate, with this ansatz we can accurately describe most of the GS properties of interest in the regime $J/U_0 \ll 1$.
    
    \section{Equilibrium Properties\label{sec:equilibrium}} For probing the accuracy of the outlined method, in this section we present a quantitative comparison of different GS observables such as energy, condensate fraction, long-range phase coherence and notably, the bi-partite entanglement entropy of the 1DBHM, computed via the the outlined methods with ED in small systems. In ED calculations, we find the lowest lying eigenstate of $\hat{H}$ for a given system size $L$, with a specified number of bosons $N_b$. The largest systems considered have Hilbert space dimension $\sim 10^5$. For the approximate calculations outlined above, we (numerically) optimize the quantum expectation value of the 1DBHM Hamiltonian, i.e. $\langle \tilde\Psi_{\text{GW}}(\boldsymbol{f})|\hat{H}|\tilde\Psi_{\text{GW}}(\boldsymbol{f})\rangle$ or $\langle \tilde\Psi_{\text{SW}}(\boldsymbol{f})|\hat{H}|\tilde\Psi_{\text{SW}}(\boldsymbol{f})\rangle$, to obtain an optimal value of $\boldsymbol{f}$, say $\boldsymbol{f}^\star$ (see Appendix-\ref{app:sec:evaluate_energy_functional} for details), and hence the approximate GS wavefunction $\ket{\tilde \Psi_{\text{GW}}(\boldsymbol{f}^\star)}$ (or $\ket{\tilde \Psi_{\text{SW}}(\boldsymbol{f}^\star)}$). We can then probe other relevant physical aspects such as condensate fraction, long-range phase coherence and bi-partite entanglement entropy of the exact GS and compare it with two different approximate ground states --- (i) particle number projected Gutzwiller state $\ket{\tilde \Psi_{\text{GW}}(\boldsymbol{f}^\star)}$ and (ii) $\ket{\tilde \Psi_{\text{SW}}(\boldsymbol{f}^\star)}$. In the following, we present our results for the homogeneous and weakly confining harmonic profiles of the chemical potential.

\begin{figure}[!htpb]
        \centering
        \rotatebox{0}{\includegraphics[width=0.238\textwidth]{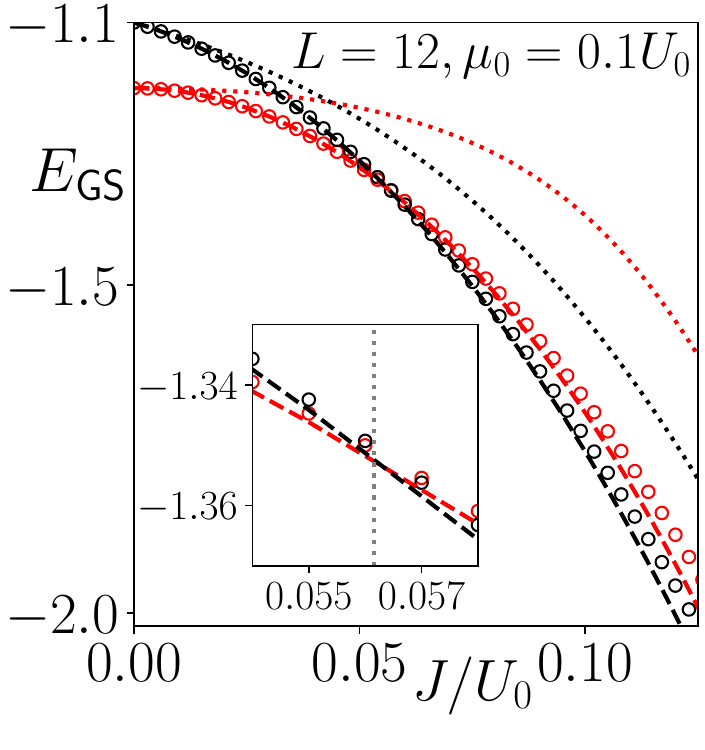}}
        \rotatebox{0}{\includegraphics[width=0.238\textwidth]{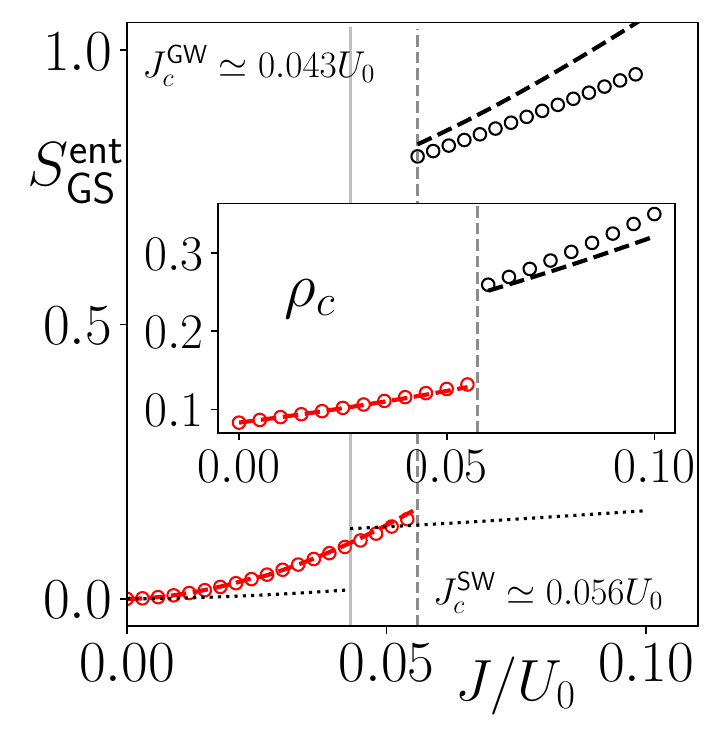}}
        \caption{ Left: Ground state energies $E_{\text{GS}}$ of for $N_b=L-1$ (black) and $N_b=L$ (red) vs $J/U_0$ using three different methods: (i) ED (open circles), (ii) GW ansatz (dotted lines) and (iii) SW ansatz (dashed lines). The energy crossing is at $J_c\sim0.056U_0$ for ED and SW ansatz while for the GW ansatz $J_c \sim 0.495U_0$. Right: Half-chain bi-partite von Neumann entanglement entropy $S_{\text{ent}}^{\text{GS}}$ and condensate fraction $\rho_c$(inset) for the same set of parameters.}
        \label{main:fig:GS_props_mu_0x1U0}
    \end{figure}

    Fig.~\ref{main:fig:GS_props_mu_0x1U0} shows the variation of the GS energy $E_{\text{GS}}$ for different $J/U_0$ obtained from (i) ED , (ii) $\langle \tilde\Psi_{\text{GW}}(\boldsymbol{f}^\star)|\hat{H}|\tilde\Psi_{\text{GW}}(\boldsymbol{f}^\star)\rangle$ and $\langle \tilde\Psi_{\text{SW}}(\boldsymbol{f}^\star)|\hat{H}|\tilde\Psi_{\text{SW}}(\boldsymbol{f}^\star)\rangle$. Due to the translational symmetry of the problem, the number of variational parameters is $N_{\text{max}}+1(=4)$. Although the total number of particles are kept fixed while obtaining the GS energies and the homogeneous chemical potential profile does not play any role other than shifting the energy eigenvalues, the different $N_b$ sectors will have different GS energies as a function of $J/U_0$. This means that, depending on the value of the chemical potential, the energy values corresponding to different $N_b$ sectors can cross each other at some value of $J/U_0 \equiv J_c/U_0$ indicating a quantum phase transition, and leading to different behavior of  GS properties on either sides of the transition. This is illustrated in Fig.~\ref{main:fig:GS_props_mu_0x1U0}(a) which shows that for $J<J_c$ and $J>J_c$, the GS is in the $N_b=L$ and $N_b=L-1$ sectors respectively, with $J_c \simeq 0.056U_0$.

   Although the ansatz is approximate, we show that it can still give an accurate description of the quantum correlations present in the system, such as the bi-partite von Neumann entanglement entropy. The bi-partite von Neumann entanglement entropy \cite{Entanglement_general_Islam2015,Entanglement_RMP_Fazio_RevModPhys.80.517} describes non-local quantum correlations between two parts of the system with respect to a given quantum state. It plays a crucial role in understanding the properties of the ground-state phases \cite{topological_order_entanglement_Jiang2012}, thermalization in isolated quantum many-body systems \cite{Entanglement_in_thermalization_doi:10.1126/science.aaf6725} and also serves as an order-parameter in the description of quantum phases \cite{Entanglement_RMP_Fazio_RevModPhys.80.517}.  Fig.~\ref{main:fig:GS_props_mu_0x1U0}(b) shows that the optimized ansatz state $\ket{\tilde\Psi_{\text{SW}}(\boldsymbol{f}^\star)}$ captures the half-chain von Neumann entanglement entropy associated with the the exact GS as a function of $J/U_0$ more accurately compared to $\ket{\tilde\Psi_{\text{GW}}(\boldsymbol{f}^\star)}$. In all cases, the GS entanglement entropy is computed using

    \begin{equation}
      S_{\text{GS}}^{\text{ent}}=-\text{Tr}_A(\rho_A\ln\rho_A)  
    \end{equation}

    with $\rho_A=\text{Tr}_B(\ket{\Psi}_{\text{GS}}\bra{\Psi}_{\text{GS}})$ and $A,B$ both spanning half of the lattice. 
    
    Another important physical quantity is the condensate fraction $\rho_c$, which is defined as
    
    \begin{equation}
        \rho_c=\frac{1}{LN_b}\sum_{r,r'=1}^L\langle \Psi_{\text{GS}}|\hat{b}_r^\dagger\hat{b}_{r'}|\Psi_{\text{GS}}\rangle
        \label{main:eq:rho_c}
    \end{equation}
    
    The inset of Fig.~\ref{main:fig:GS_props_mu_0x1U0}(b) shows $\rho_c$ for the same parameter choices as in the main panel. While $\rho_c$ exhibits a jump at $J=J_c$ in all three methods, the SW ansatz shows quantitatively better agreement with ED compared to GW ansatz.

    Furthermore, the state $\ket{\Psi_{\text{SW}}(\boldsymbol{f}^\star)}$ captures the correlations between different sites of the lattice and thus can distinguish between different GS phases based on the nature of this correlation. To see this we consider the single particle Green's function which is defined as the correlation function 

    \begin{equation}
        C(r)=\langle \Psi_{\text{GS}}|\hat{b}_0^\dagger\hat{b}_r|\Psi_{\text{GS}}\rangle   
    \end{equation}    
    
\begin{figure}[!htpb]
        \centering
        \rotatebox{0}{\includegraphics[width=0.49\textwidth]{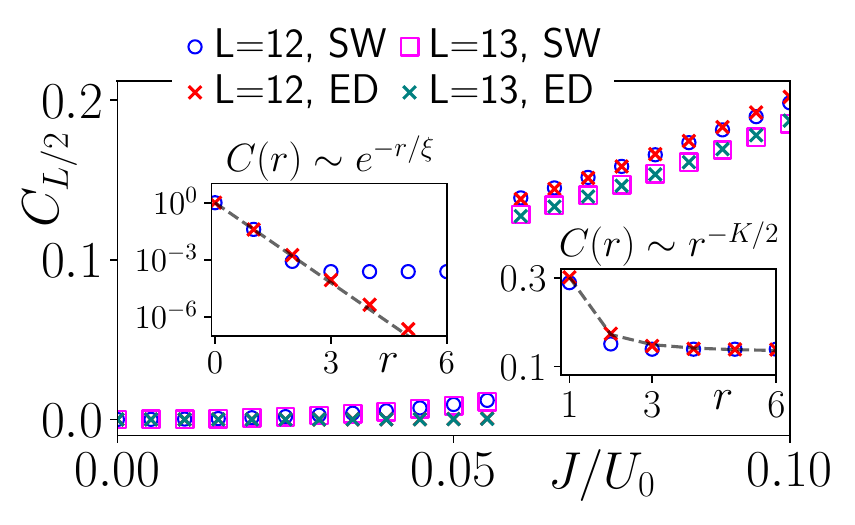}}
        \caption{Behavior of long-range phase coherence $C(L/2)$ vs $J$ (main panel) and the different natures of the decay of the correlation function $C(r)$ as a function of $r$ from an exponential at $J=0.01U_0$ (left inset) to a power-law at $J=0.2U_0$ (right inset)}
        \label{main:fig:GS_props_ODLRO_mu_0x1U0}
    \end{figure}

   In the main panel of Fig.~\ref{main:fig:GS_props_ODLRO_mu_0x1U0} we show the variation of $C(L/2)$, which is a measure of long-range phase coherence \cite{Yang_RevModPhys.34.694} on a finite lattice, as a function of $J/U_0$ using ED (red crosses) and using the approximate GS $\ket{\tilde \Psi_{\text{SW}}(\boldsymbol{f}^\star)}$ (blue circles) for two different system sizes $L=12,13$ and $N_b=L,L-1$. The value of $C(L/2)$ decreases for higher system size and signifies the absence of true off-diagonal long-range ordering in the thermodynamic limit for the 1DBHM. Furthermore, the insets shows that this correlation function decays exponentially ($C(r) \sim e^{-r/\xi}$) for $J/U_0 < J_c/U_0$ while it shows a power-law decay ($C(r) \sim r^{-\alpha}$) in the regime $J/U_0 > J_c/U_0$  as a function of distance $r$ on the lattice. 
    
    The variation of $C(r)$ obtained from $\ket{\tilde \Psi_{\text{SW}}(\boldsymbol{f}^\star)}$ (blue circles) matches with that obtained from the exact calculations (red crosses) up to $r=3$, after which it flattens out. This behavior can be attributed to the finite order truncation of the perturbative SW transformation. \\

    \begin{figure}[!htpb]
        \centering
        \rotatebox{0}{\includegraphics[width=0.48\textwidth]{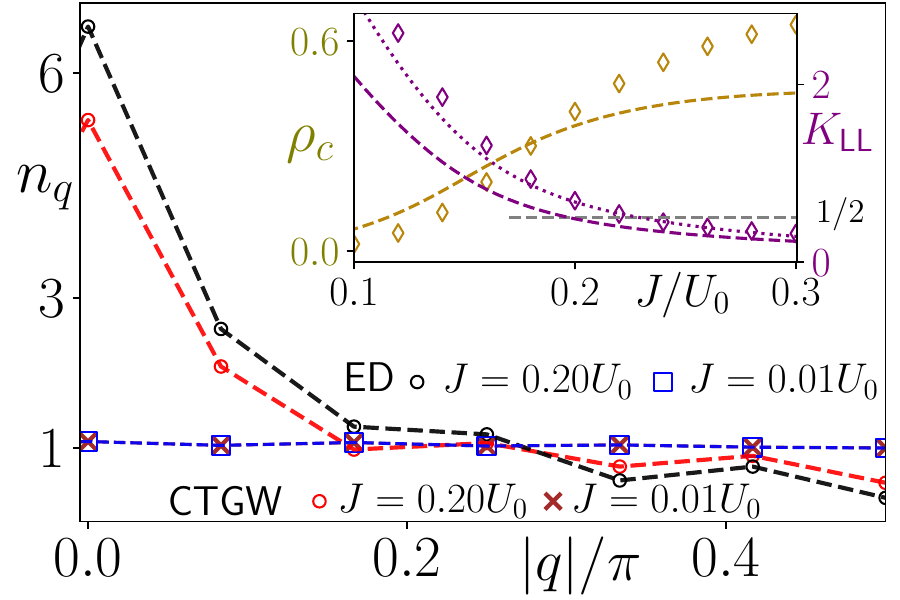}}
        \caption{ Distribution of quasi-momentum population $n_q$ with for $N_b=L$ and $J=0.01U_0,0.2U_0$. The enhancement of $n_{q=0}$ for $J=0.2U_0$, as well as the continuous growth of $\rho_c$ with $J$ (in the inset ; yellow diamond points for ED and yellow dashed line from the approximate ground state) indicates the possible presence of a BKT transition at integer filling being captured by the approximate GS. In the inset, shows $K_{\text{LL}}$ with $J$ across the transition at integer filling. See text for details.}
        \label{main:fig:GS_props_BKT}
    \end{figure}
    
    Moreover, the study of the 1DBHM in the canonical ensemble provides a simple and direct framework to analyze how much we can infer about the Berezinskii-Kosterlitz-Thouless (BKT)-type transition which occurs in the 1DBHM for fixed densities \cite{Kuhner_White_Monien_PhysRevB.61.12474}, using the approximate GS $\ket{\tilde\Psi_{\text{SW}}(\boldsymbol{f}^\star)}$. To this end, we consider the off-diagonal matrix elements of the one-body density matrix ($\langle \Psi_{\text{GS}} | \hat{b}_i^\dagger \hat{b}_j | \Psi_{\text{GS}} \rangle$) as a function of $J/U_0$. The correlation length exponent $\alpha$ is related to the so-called Luttinger parameter $K_{\text{LL}}$ as $\alpha = K_{\text{LL}}/2$. This exponent can be obtained by fitting $C(r) \sim r^{-\alpha}$ and at the BKT transition point, one has $\alpha=1/4,K_{\text{LL}}=1/2$ \cite{Kuhner_White_Monien_PhysRevB.61.12474}. In the inset of Fig.~\ref{main:fig:GS_props_BKT} we show the variation of $K_{\text{LL}}$ obtained from the exact GS (purple points). For the SW ansatz we adapt two different fitting procedures to obtain $K_{\text{LL}}$: we fit $C(r)\sim r^{-\alpha}$ in the range $r=0-3$ (purple dotted) and also in the range $r=0-L/2$ (purple dashed lines). The dotted lines show better alignment with the exact answer.
    As shown in the inset in Fig.~\ref{main:fig:GS_props_BKT}, the approximate SW ansatz state also predicts a continuous growth of $\rho_c$ for $N_b=L$ in line with the ED calculations. 
 
    Using interferometry techniques \cite{Bloch_nature_2010}, it is possible to construct the momentum space distribution of the occupation number and it is an important diagnostic observable in experiments. In the main panel of Fig.~\ref{main:fig:GS_props_BKT} we present a quantitative comparison of the quasi-momentum distribution $n_q$ defined as, 
    
    \begin{equation}
        n_q = \frac{1}{L}\sum_{r,r'=1}^L\langle \Psi_{\text{GS}}|\hat{b}_r^\dagger\hat{b}_{r'}|\Psi_{\text{GS}}\rangle e^{i q (r-r')}
    \end{equation}

    obtained using ED and $\ket{\tilde \Psi_{\text{SW}}(\boldsymbol{f}^\star)}$ at integer filling in the $N_b=L$ sector for $J=0.01U_0$ and $J=0.2U_0$ respectively.\newline

\begin{figure}[!htpb]
        \centering
        \rotatebox{0}{\includegraphics[width=0.48\textwidth]{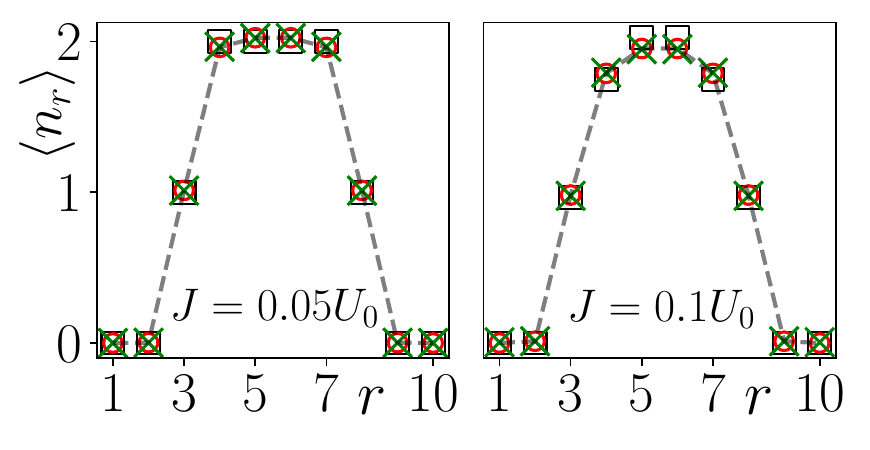}}
        \caption{Average occupation-number profile $\langle n_r \rangle$ over the lattice for $J=0.05U_0$ (left panel) and $J=0.1U_0$ (right panel), for $L=N_b=10,\mu_0=0.2U_0,\kappa=0.345U_0$. Different point types indicate the different optimized approximate GS (\textit{green cross} for SW and \textit{black square} for GW) and exact GS (\textit{red circles}). The dashed gray line is a guide to the eye.}
        \label{main:fig:GS_trap}
    \end{figure}

    We now consider the effect of a weak harmonic confining potential $ \mu_r = \mu_0 - \frac{1}{2}\kappa(r-r_0)^2$, with $r_0,\kappa$ being the center of the lattice and the strength of trapping respectively. This allows us to demonstrate the applicability of the method for finding GS properties beyond the homogeneous chemical profiles discussed earlier. Such  trapping potentials are always present in experiments with ultracold atomic gases \cite{Bloch_Dalibard_Zwerger_RevModPhys.80.885}. As the chemical potential $\mu_r$ is now site-dependent, the Gutzwiller parameters needed to write down the appropriate ansatz state are also site-dependent, i.e. $\boldsymbol{f}\equiv \{ f_r^{n_r} \}$, and the total number of variational parameters is thus $L(N_{\text{max}}+1)$. Although it is possible to make the number of variational parameters $(N_{\text{max}}+1)L/2$ by taking into account the reflection symmetry of the confining potential about it's center, we have not made an explicit use of this during the optimization. Fig.~\ref{main:fig:GS_trap} shows the average on-site occupation numbers for the GS of 1DBHM for $L=10,N_b=10$ in presence of a harmonic confining potential with $\kappa=0.345 U_0$. As depicted in this figure, the optimized state $\ket{\tilde\Psi_{\text{SW}}(\boldsymbol{f}^\star)}$ (green cross) shows good agreement with the exact answer (red circles).   
    
    \section{Dynamics\label{main:sec:dynamics}} 
    
    We now study how accurately various ansatz states introduced earlier, or variants of them, can capture non-equilibrium dynamics of the 1DBHM. We shall consider three scenarios: (i) dynamics of a BEC-like state in a deep optical lattice (Sec.~\ref{main:subsec:BEClikestate-dynamics}), (ii) dynamics in a shifted harmonic trap (Sec.~\ref{main:subsec:dyn_shifted_trap}) and (iii) Quench dynamics from a family of $U(1)$ symmetry broken states (Sec.~\ref{main:subsec:QME}). For scenarios (i),(ii) we will take $\ket{\tilde \Psi_{\text{GW}}(\boldsymbol{f})}$ as a variational state and evolve the equations of motion for the parameters ${\boldsymbol{f}}$ obtained using TDVP to approximate the quantum evolution. For scenario (iii) we take the state $e^{gi\hat{S}}\ket{\tilde \Psi_{\text{GW}}(\boldsymbol{f})}$ as the state and $\{\boldsymbol{f},g\}$ as the variational parameters. 
    
    Before going to the details of specific scenarios, we briefly describe our approach to approximate the quantum dynamics within a chosen variational manifold. We will apply the time-dependent variational principle (TDVP) \cite{Kramer_ref1,10.21468/SciPostPhys.9.4.048_Hackl_Cirac} to a specific variational ansatz, say $\ket{\Psi(\boldsymbol{x})}$, to construct optimal equations of motion for the parameters, $\boldsymbol{x}$. We chose to work with real parameters $\boldsymbol{x}\in\mathbb{R}^M$, where $M$ is the number of variational parameters. In cases where the state requires some complex parameters, we treat the real and imaginary parts of the parameters as independent real parameters. If the state $\ket{\Psi(\boldsymbol{x})}$ is normalized, i.e. $\langle \Psi(\boldsymbol{x})|\Psi(\boldsymbol{x})\rangle=1$, then the time-dependent variational principle amounts to extremizing the action $\mathcal{S}=\int dt \; \mathcal{L}$, where the Lagrangian reads $\mathcal{L}=\mathfrak{Re}(\langle \Psi(\boldsymbol{x}) |i\frac{d}{dt} -\hat{H}|\Psi(\boldsymbol{x})\rangle)$. This gives rise to the following Euler-Lagrange equations of motion

    \begin{equation}
       \sum_{b=1}^M \mathcal{G}_{ab}(\boldsymbol{x}(t)) \; \dot{\boldsymbol{x}}_b(t) = \mathcal{F}_a(\boldsymbol{x}(t))
        \label{main:eq:TDVP-EOMs}
    \end{equation}
    
    for the parameters $\boldsymbol{x}(t)$, with $\mathcal{G}_{ab}(\boldsymbol{x})$, $\mathcal{F}_a(\boldsymbol{x})$ for $a,b=1,2,...,M$ given as

    \begin{subequations}
    \begin{align}
        \mathcal{G}_{ab}(\boldsymbol{x}) & = 2 \mathfrak{Im}\left[\langle \partial_a \Psi(\boldsymbol{x}) | \partial_b \Psi(\boldsymbol{x}) \rangle\right] \\
        \mathcal{F}_{a}(\boldsymbol{x}) & = -2 \mathfrak{Re}\left[\langle \partial_a \Psi(\boldsymbol{x}) |\hat{H}|\Psi(\boldsymbol{x}) \rangle \right]
        \label{eq:Gram_matrix_force_vector}
    \end{align}
    \end{subequations}
    
    These equations are a set of coupled, first-order, non-linear ordinary differential equations and will be henceforth referred to as the TDVP equations of motion. To solve these equations it is necessary to formally invert the matrix $\mathcal{G}(\boldsymbol{x})$ and write  $\dot{\boldsymbol{x}}_a(t) = \mathcal{G}^{-1}_{ab}(\boldsymbol{x}) \mathcal{F}_b(\boldsymbol{x})$.  Now, the key computational task remaining is to construct $\mathcal{G}(\boldsymbol{x}),\mathcal{F}(\boldsymbol{x})$ numerically, which can be done following the prescription outlined in Appendix-\ref{app:sec:TDVP_EOM_construction}. We now discuss the three scenarios alluded to earlier. 

    \subsection{BEC-like state in a deep lattice: \label{main:subsec:BEClikestate-dynamics}} The ideal BEC state for a fixed number bosons $N_b$ reads \cite{Zoller_atoic_matter_wave_revival_PhysRevA.83.043614}

            \begin{equation}
                \ket{\text{BEC}(N_b)} \equiv \sqrt{\frac{N_b!}{N_b^{N_b}}} \sideset{}{'} \sum_{\{n\}} \frac{1}{\sqrt{n_1!n_2!...n_L!}} \ket{n_1,n_2,...,n_L}
                \label{main:eq:BECstate_with_Nb_bosons}
            \end{equation}

\begin{figure}[!htpb]
        \centering
        \includegraphics[width=0.48\textwidth]{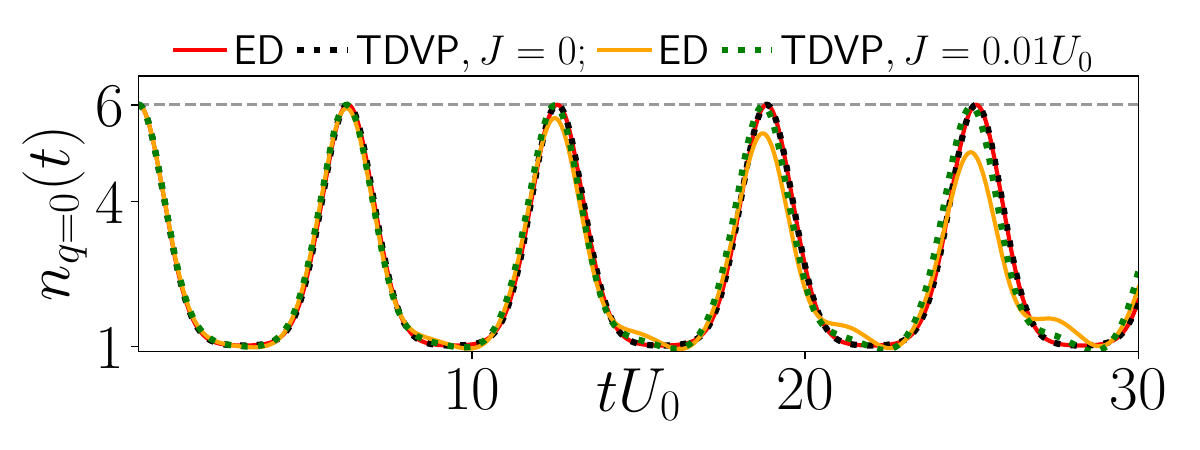}
        \caption{Time-evolution of $n_{q=0}$ starting from a BEC-like state using TDVP (dotted lines) and ED (solid lines) in two different quench scenarios (i) $J=0$ and (ii) $J=0.01U_0$. The figure is for $L=10,N_b=10$.}
        \label{main:fig:BEC_quench}
    \end{figure}

    In the above expression the symbol $\sideset{}{'}\sum_{\{n\}}$ denotes summation over all occupation number configurations $\{n\}$ which satisfy $\sum_r n_r=N_b$. Since our variational ansatz has a different form compared to Eq.~\eqref{main:eq:BECstate_with_Nb_bosons}, in order to mimic the BEC state we minimize $||\ket{\tilde \Psi_{\text{GW}}(\boldsymbol{f})}-\ket{\text{BEC}(N_b)}||^2$ with respect to $\boldsymbol{f}$ and find the Gutzwiller parameters, say $\boldsymbol{f}_{\text{BEC}}$, which best emulate the $\ket{\text{BEC}(N_b)}$ state from within our variational ansatz. We study the dynamics of $n_{q=0}$ i.e.,

    \begin{equation}
        n_{q=0}(t) = \frac{1}{L}\sum_{r,r'=1}^L\langle \tilde\Psi_{\text{GW}}(t)|\hat{b}_r^\dagger\hat{b}_{r'}|\tilde \Psi_{\text{GW}}(t)\rangle
    \end{equation}
    
    starting from the state $\ket{\tilde \Psi_{\text{GW}}(\boldsymbol{f}_{\text{BEC}})}$ under a quench of optical lattice depth. As shown in the Fig.~\ref{main:fig:BEC_quench} it shows perfect revival when the lattice is infinitely deep (implying $J/U_0=0$) and this oscillations are eventually washed away when the depth of the lattice is finite. The time dependence of variational parameters can be obtained analytically in an infinitely deep ($J/U_0=0$) lattice and is given by $f^n(t) = f^n(0) \exp\left( -i t (-\mu_0 n + U_0n(n-1)/2 \right)$. For a finite depth one needs to evolve the TDVP equations of motion as described by Eq.~\eqref{main:eq:TDVP-EOMs} or evolve the system using ED. As depicted in Fig.~\ref{main:fig:BEC_quench}, the rate of decay of these oscillations is very slow within the TDVP framework, and it only captures the exact evolution for short time-scales (i.e. up to $tU_0 \sim 20$). 
    
    \subsection{Quench in a Harmonic Confinement \label{main:subsec:dyn_shifted_trap}} In order to emphasize the applicability of the TDVP principle on particle-number projected Gutzwiller-like states in the presence of inhomogeneities, we study quantum quench dynamics in the 1DBHM in presence of a harmonic trapping potential. We start with an approximate GS of $\hat{H}$ and then suddenly change some parameters of the Hamiltonian. In particular, we study two scenarios: (i) quenching of hopping strength $J$ from $J_i$ to $J_f$ in the presence of an external harmonic trapping and  (ii) instantaneous shifting of the center of the external harmonic trapping potential while keeping hopping strength $J$ unchanged. 
    
\begin{figure}[!htpb]
        \centering
        \includegraphics[width=0.45\textwidth]{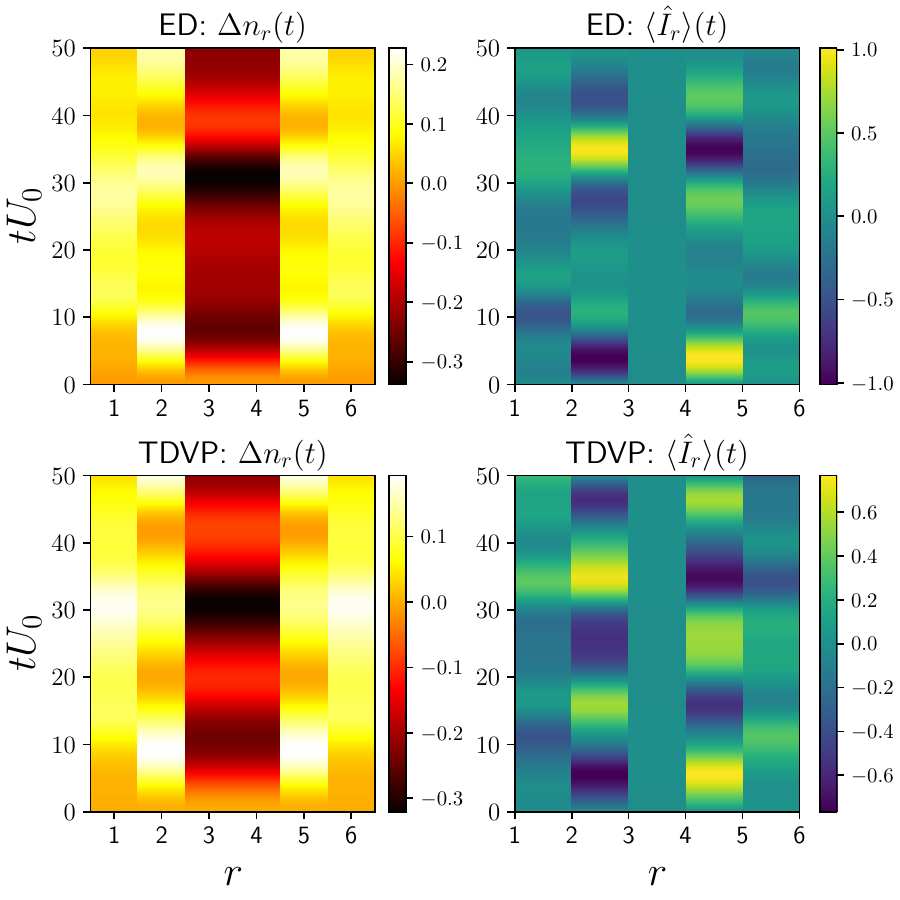}
        \includegraphics[width=0.48\textwidth]{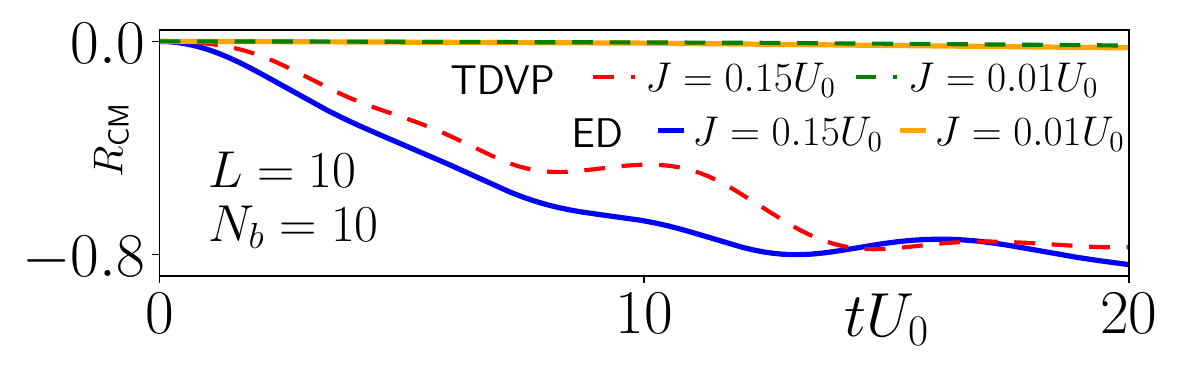}
        \caption{Quench Dynamics: \textit{top and middle panels} -- Comparison of the time dependence of site-resolved occupation number change $\Delta n_r(t)$ and local particle current $\langle\hat{I}_r(t)\rangle$ with $J_i=0.01U_0$,$J_f=0.05U_0$ using ED (top panels) and TDVP (middle panels). \textit{bottom panel}: Comparison of the time evolution of $R_\text{CM}(t)$ under the quench of the center of the trap position by one lattice unit -- using ED (solid lines) and TDVP (dashed lines) for $J/U_0=0.01,0.2$.  For all panels, $\kappa=0.345U_0$.}
        \label{fig:quench_trap}
    \end{figure}

    For scenario (i) we start with the GS of the system at $J=J_i$ of Eq.~\eqref{eq:BHmodel-1D} ($\ket{\tilde\Psi_{\text{GW}}(\boldsymbol{f}=\boldsymbol{f}^\ast)}$ where ${\boldsymbol{f}^\ast}$ are Gutzwiller parameters obtained by optimizing the expectation value of canonically transformed effective Hamiltonian with respect to $\ket{\tilde\Psi_{\text{GW}}(\boldsymbol{f})}$) and perform a sudden quench of the hopping strength $J$ to $J_f$. We evolve the quantum state using (i) TDVP equations of motion obtained from the $\ket{\tilde \Psi_{\text{GW}}(\boldsymbol{f}(t))}$ and (ii) ED method. As it is evident from the Fig.~\ref{fig:quench_trap} the site-resolved occupation number change, 

    \begin{equation}
      \Delta n_r(t) = \langle \hat{n}_r \rangle(t)-\langle \hat{n}_r \rangle(0)  
    \end{equation}
    
    and expectation values of the local currents,

    \begin{equation}
        \hat{I}_r(t)= \frac{1}{L}{\sum_{\langle r' \rangle_r} (\hat{b}^\dagger_r \hat{b}_{r'} -\hat{b}^\dagger_{r'} \hat{b}_r)}
    \end{equation}
    
    obtained from exact evolution are well described via the TDVP evolution of the particle-number projected Gutzwiller variational state. In the expression above the sum on $r'$ is over all first nearest neighbors of the site $r$, and is symbolically denoted by $\langle r' \rangle_r $

    For scenario (ii) we start with the approximate GS of the system at $J=J_i$ and shift the external harmonic trapping potential by one lattice unit. This has the effect of an instantaneous quenching of the external harmonic trapping, i.e. $\mu_r(t) = \mu_0 - \frac{1}{2}\kappa(r-R(t))^2$ with $R(t=0)=r_0$ and $R(t>0)=r_0+1$, $r_0$ being the center of the lattice. As expected, it is evident from the bottom panel of  Fig.~\ref{fig:quench_trap} that for the values of $J_i$ which correspond to a deep Mott-insulator regime, the center of mass of the ensemble of bosonic particles

    \begin{equation}
        R_{\text{CM}}(t) \equiv \frac{1}{N_b} \sum_{r=1}^L {r\langle \hat{n}_r \rangle(t) }
    \end{equation}
    
    is almost frozen and TDVP dynamics gives a very accurate description of the exact evolution. On the other hand larger values of $J_i$, results in a faster relaxation of $R_{\text{CM}}$ which is also captured qualitatively  within the framework of TDVP. The TDVP dynamics is performed by solving the TDVP equations of motion with $(N_{\text{max}}+1) L$ Gutzwiller parameters. Here, $N_{\text{max}}=3$ is the finite cutoff introduced for the maximum number of allowed bosons per site in the strongly correlated regime. As this number of parameters in this approximate method scales only linearly with the system size, this method is promising in capturing the dynamical signatures in a large class of strongly correlated systems even in the presence of inhomogeneities. \newline

    \subsection{Quantum Mpemba Effect \label{main:subsec:QME}} The quantum Mpemba effect (QME) is an example of an anomalous relaxation in the non-equilibrium dynamics of quantum systems, where initial states further away from equilibrium are found to relax faster compared to initial states closer to the equilibrium \cite{cala1,cala2,cala3,cala4,cala5,cala6,cala7,cala8,cala9,cala10}. We show here that a QME occurs in the 1DBHM starting from a class of initial states that explicitly break the global $U(1)$ symmetry $\hat{N}$ of $\hat{H}$ and this state is evolved under the Hamiltonian Eq.~\eqref{eq:BHmodel-1D} which respects this $U(1)$ symmetry. We show that such an evolution can be qualitatively captured within the paradigm of TDVP by considering a variational state of the form $\ket{\tilde\Psi_V(\boldsymbol{x})} = e^{g(t)i\hat{S}}\ket{\tilde\Psi_{\text{GW}}(\boldsymbol{f}(t))}$. The specific class of initial states we work with are 

    \begin{equation}
        \ket{\psi(\theta,\phi)} = \bigotimes_{r=1}^L \left( 
       f^0_{\theta,\phi}\ket{0}_r + f^1_{\theta,\phi} \ket{1}_r + f^2_{\theta,\phi} \ket{2}_r \right)
        \label{eq:MpembaEffectStateDefn}
    \end{equation}

    where $f^0_{\theta,\phi}=\cos(\theta),f^1_{\theta,\phi}=\sin(\theta)\cos(\phi),f^2_{\theta,\phi}=\sin(\theta) \sin(\phi)$, $\theta\in[0,\pi],\phi\in[0,2\pi]$. We note here that for this family of initial states $g(0)=0$. Since $\hat{N}$ is a symmetry of $\hat{H}$, it is expected that time evolution of this state under $\hat{H}$, dynamically restores the $U(1)$ symmetry at the level of sub-system. We have studied the behavior of two quantities, the sub-system entanglement asymmetry \cite{cala1} (see Eq.~\eqref{main:eq:subsytem-ent-asymm}) and sub-system evolution speed \cite{Khasseeh_Heyl_Jiaju_Rajabour_PhysRevB.111.L140410} (see Eq.~\eqref{main:eq:subsytem-evol-speed}) to detect the presence of a QME. The first probe directly measures the degree of symmetry breaking in the chosen initial state for a particular choice of the subsystem, and its time evolution concerns the dynamical restoration of that symmetry if the corresponding Hamiltonian that governs the evolution of the system respects that particular symmetry broken by the initial state. On the other hand, the second probe quantifies the instantaneous speed of evolution of the reduced density matrix of the sub-system of interest. Although the second probe does not directly concern the global symmetry of the system or require any symmetry resolution, it is instructive to look at both measures for a symmetry broken initial state. The entanglement asymmetry of a subsystem $\Delta S_{\text{ent}}^A$ of a (pure) quantum state $\ket{\psi}$ is defined as

    \begin{equation}
        \Delta S_{\text{ent}}^A = S_{\text{ent}}(\hat{\rho}_{A,N}) - S_{\text{ent}}(\hat{\rho}_{A})
        \label{main:eq:subsytem-ent-asymm}
    \end{equation}

     where $S_{\text{ent}}(\rho_A)=-\text{Tr}_A\left(\hat{\rho}_A\ln\hat{\rho}_A\right)$, $\hat{\rho}_A=\text{Tr}_B\left(\ket{\psi}\bra{\psi}\right)$ is the reduced density matrix (for subsystem $A$, $B$ denotes the rest of the system) of the pure quantum state  $\ket{\psi}$ and $\hat{\rho}_{A,N}=\sum_{n_b} \hat{\Pi}_{n_b} \hat{\rho}_A \hat{\Pi}_{n_b}$, with $\hat{\Pi}_{n_b}$ being the projection operator onto the sector with $n_b$ particles in the subsystem $A$. For evaluating a quantity such as $\Delta S_{\text{ent}}^A$, it is imperative that the $U(1)$ symmetry $\hat{N}$ of $\hat{H}$ is maintained explicitly while performing the time evolution of the initial state $\ket{\psi(\theta,\phi)}$, something which cannot be achieved within a grand-canonical ensemble. Since the ansatz $\ket{\tilde\Psi_V(\boldsymbol{x})}$ works directly within the canonical ensemble, it is possible to decompose a state $\ket{\psi_{\text{mixed}-N_b}}$, which is a superposition of different total particle-number sectors, mathematically as

     \begin{equation}
         \ket{\psi_{\text{mixed}-N_b}} = \bigoplus_{n_b \in N_b^{\text{sectors}}} \ket{\tilde \Psi^{(n_b)}_{V}(\boldsymbol{x}^{(n_b)})}
     \end{equation}

    The approximate time-evolution of the initial state $\ket{\psi(\theta,\phi)}$ then amounts to evolving the TDVP equations of motions for the parameters $\boldsymbol{x}^{(n_b)}(t)$ for all allowed sectors $n_b \in N_b^{\text{sectors}}$. \newline 
    
    \begin{figure}[!htpp]
        \centering
        \includegraphics[width=0.48\textwidth]{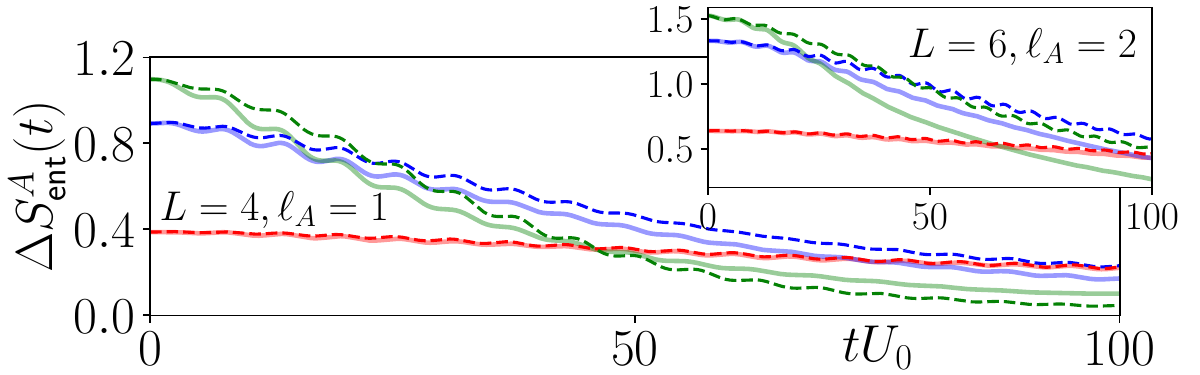}
        \includegraphics[width=0.48\textwidth]{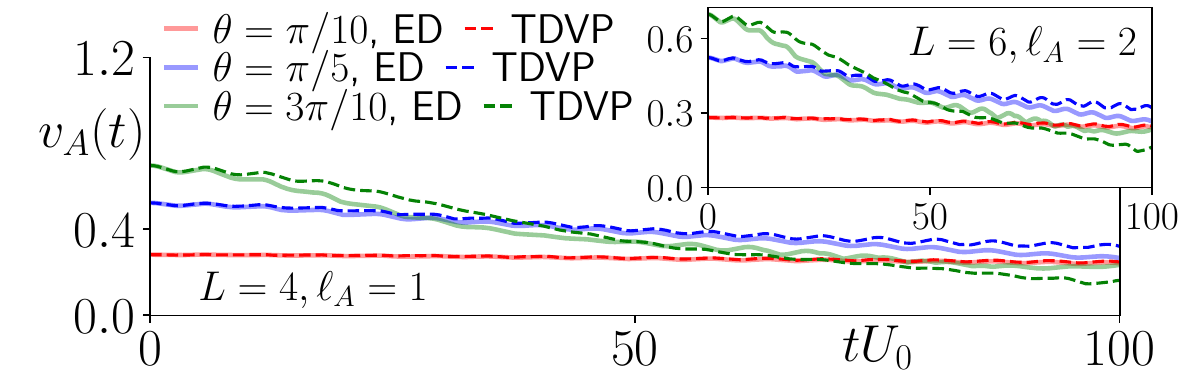}
        \caption{Quantum Mpemba effect in the 1DBHM: Relaxation dynamics of three different initial states ($\theta=\pi/10,\pi/5,3\pi/10$ with $\phi=\pi/4$) observed via (a) entanglement asymmetry $\Delta S_{\text{ent}}^A(t)$ and (b) sub-system evolution speed $v_A(t)$ for $L=4,\ell_A=1$ and $L=6,\ell_A=2$ (insets) using ED (dashed curves) and TDVP (solid curves)}
        \label{fig:QME}
    \end{figure}    
    
    As Fig.~\ref{fig:QME}(a) illustrates, our numerical calculations indicate the presence of a QME in the setup described above. Starting from three different initial states ($\theta=\pi/10,\pi/5,3\pi/10$ and $\phi=\pi/4$) computed using TDVP and ED, we find that the state corresponding to the highest value of $\Delta S_{\text{ent}}^A(t)$ relaxes faster compared to the other two initial states, and these curves cross, implying the onset of a QME.
    
    Since, the Hamiltonian $\hat{H}$ under which the system evolves has a global $U(1)$ symmetry, the presence of a QME could be detected using the entanglement asymmetry. We now explore the second alternative probe --- the sub-system evolution speed --- and show that it can also detect the presence of a QME, without appealing to any global symmetries of $\hat{H}$. The sub-system evolution speed $v_A(t)$ is defined as \cite{Khasseeh_Heyl_Jiaju_Rajabour_PhysRevB.111.L140410}

    \begin{equation}
        v_A(t) = \lim_{\delta t\rightarrow 0} \frac{D(\hat{\rho}_A(t+\delta t),\hat{\rho}_A(t))}{\delta t}
        \label{main:eq:subsytem-evol-speed}
    \end{equation}    

     which quantifies the rate at which the state of sub-system $A$ changes in terms of it's instantaneous reduced density matrix $\rho_A(t)$, and $D(\hat{\rho}_1,\hat{\rho}_2)$ is suitable a distance function between two (density) matrices $\hat{\rho}_{1,2}$. For the numerical results shown, we have chosen the trace norm distance, $D(\hat{\rho}_1,\hat{\rho}_2)=|\!|\hat{\rho}_1-\hat{\rho}_2|\!|_1$, where $|\!| . |\!|_1$ denotes the trace norm; although other choices are also valid (see \cite{Khasseeh_Heyl_Jiaju_Rajabour_PhysRevB.111.L140410}). 

     By looking at both panels of Fig.~\ref{fig:QME} we can conclude that for $\theta=3\pi/10$, the initial state has a higher degree of symmetry breaking and also the rate at which the sub-system $A$ is evolving is much faster for this state compared to the other states. As a result of this higher speed, the subsystem is able to relax much faster compared to the other initial states, and the entanglement asymmetry also decreases much faster, leading to the QME. Thus the subsystem evolution speed plays qualitatively the same diagnostic role as quasi-particle velocities in integrable systems exhibiting QME \cite{cala1}. Since the concept of a sub-system evolution speed is very general, it could be a useful indicator for detecting QME in systems where one does not have any global symmetries and/or the system under consideration in non-integrable and one does not have a simple quasi-particle picture to understand the relaxation processes. Recent works \cite{bhore2025quantummpembaeffectglobal} have investigated QME is non-integrable systems devoid of global symmetries by explicitly computing the distance of the instantaneous state from the expected late-time steady state. Using sub-system velocities in this context may also be helpful as this gives us an estimate of the rate of instantaneous evolution, even when the nature of the late-time steady state is not known precisely.

    \section{Discussion and Outlook \label{sec:discussion}} In summary, the central idea of this paper is to project the Gutzwiller (GW) variational state onto a fixed particle-number sector (in the same spirit as \cite{Number_projected_Gutz_Krauth_PhysRevB.45.3137,Rokhshar_generalised_GW_PhysRevB.44.10328}). This serves two purposes: (i) it manifestly respects the global conservation of the total number of particles, and (ii) such a projection introduces non-trivial correlations in the Gutzwiller state, and as a consequence, it now has non-zero entanglement. It was shown that supplementing this projected Gutzwiller variational state with an additional projection onto an effective low-energy manifold (if a separation of energy scales exists) results in faithful reproduction the ground-state properties. For the 1D Bose-Hubbard model with strong on-site repulsive interactions, such an effective low-energy manifold can be readily constructed, for example, by designing a perturbative canonical (Schrieffer-Wolff) transformation. As a consequence of this, it becomes possible to accurately describe the physical properties using a Schrieffer-Wolff rotated particle-number projected Gutzwiller-like ansatz state. After constructing this ansatz, the problem becomes a multi-dimensional optimization problem with the expectation value of the Hamiltonian with respect to this newly designed ansatz state as the cost function and the Gutzwiller weights as the variational parameters. Since this method can capture quantum correlations, and also incorporates all other conservation laws of the system, (such as particle-number-conservation and translational symmetry), such approaches opens up a new path towards the exploration of a large class of phenomena in ultracold atomic platforms. \\

     Furthermore, we have shown that a time-dependent variational principle (TDVP) \cite{Haegeman_PhysRevLett.107.070601} based on this projected and the Schrieffer-Wolff rotated Gutzwiller ansatz state, one can also capture the non-equilibrium dynamics of such systems. Particularly, we have shown that solving the equations of motion of these few variational parameters within the TDVP framework, it becomes possible to capture the essential features of the short-time quench dynamics from a shallow to a deep optical lattice. We have also demonstrated how such semi-classical TDVP equations of motion of the variational parameters are able to capture the short-time relaxation dynamics in the presence of a trapping potential. The application of the TDVP principle is not expected to give accurate results for an arbitrary initial state, meaning the applicability of this approach is restricted. That being said, by working within a given variational manifold of ansatz states, we have been able to identify a class of interesting initial states, which break the global U(1) symmetry and exhibits the quantum Mpemba effect and this is within the reach of the TDVP approach.

     There are several possible future directions can be explored within this framework. Firstly, it is quite intuitive that, this approach is easily generalizable to a larger class of Hamiltonians containing different mixtures of bosonic species. The method introduced relies on two projections, one of them being the projection on a fixed particle-number sector (in case of $U(1)$ conserving systems) and the other being the projection on an effective low-energy manifold \cite{KS_twosp_PhysRevB.72.184507}. By looking at the effects of these individual projections on the ground-state properties of the system, one can develop an understanding of the physical processes playing an important role, which are crucial to capture the essential properties of the exact ground-state of the system. This is a useful exercise even for systems of relatively small system sizes. Lastly, as the experimental platforms are prone to various forms of atomic loss processes \cite{Zoller_open_BHM_PhysRevLett.105.015702} and decoherence mechanisms, it would be interesting to study if the approaches outlined in this paper would be of use in the study of environmental effects \cite{Zoller_open_BHM_PhysRevLett.105.015702,Zoller_spontaneous_emission_PhysRevA.82.063605} in such optical lattice platforms.\newline

    \section*{Acknowledgments}

   The author would like to thank Krishnendu Sengupta, Masahito Ueda and Peter Zoller for their helpful suggestions and comments. The author thanks Mainak Pal for numerous stimulating discussions.

\begin{appendix}
\numberwithin{equation}{section}

    \section{Canonical transformation and effective low-energy Hamiltonian \label{app:sec:SWrotHeff_derivation}}

        In this section, we provide a derivation of the expressions for $i\hat{S}$ and $\hat{H}_{\text{eff}}$ given in the main text. In order to construct the generator that removes the hopping processes which cost an energy of $\mathcal{O}(U_0)$ let us consider the following decomposition of the Hamiltonian of the Bose-Hubbard model

        \begin{align}
            \hat{H} & = \hat{H}_0 + \hat{T} \label{app:eq:Htot} \\
            \hat{H}_0 & = \sum_i -\mu_i\hat{n}_i + \frac{U_0}{2}\hat{n}_i(\hat{n}_i-1)  \label{app:eq:Hdiag} \\
            \hat{T} & = -J \sum_{\langle ij \rangle} \big(\hat{b}_i^{\dagger}\hat{b}_{j} + \hat{b}_j^{\dagger}\hat{b}_{i} \big)
            \label{app:eq:Hoffdiag}
        \end{align}

        Here $\hat{H}_0$ is the diagonal part in the occupation number basis and $\hat{T}$ is the off-diagonal hopping part. One can identify the low- (i.e. $\mathcal{O}(J)$) and high-energy (i.e. $\mathcal{O}(U_0)$) hopping processes by rewriting the matrix elements of the off-diagonal part $\hat{T}$ in the occupation number (Fock) basis as follows (see Ref.~\cite{KS_CT_AD_PhysRevB.86.085140})
        
        \begin{equation}
            \langle {\{m\}}|\hat{T}|\{n\} \rangle = \langle {\{m\}}| \sum_{\langle ij \rangle, \ell} \left( \hat{T}_{ij\ell}^{\alpha_{ij}=\alpha_{ij}^0} +  \hat{T}_{ij\ell}^{\alpha_{ij} \ne \alpha_{ij}^0} \right) |\{n\} \rangle
        \end{equation}

        In the above expression, $\hat{T}_0 \equiv \sum_{\langle ij \rangle, \ell}\hat{T}_{ij\ell}^{\alpha_{ij}=\alpha_{ij}^0}$ and $\hat{T}_1 \equiv \sum_{\langle ij \rangle, \ell}\hat{T}_{ij\ell}^{\alpha_{ij}\ne\alpha_{ij}^0}$ are the low- and high-energy parts of the hopping respectively. The action of the operator $\hat{T}_{ij\ell}^{\alpha_{ij}}$ in the occupation basis has the following form
        
        \begin{subequations}
            \label{app:eq:Tij12}       
            \begin{align}
                \hat{T}_{ij1}^{\alpha_{ij}} \ket{\{n\}} = -J \;\hat{b}_i^{\dagger}\hat{b}_j \;\delta(n_i-n_j-\alpha_{ij}+1) \ket{\{n\}} \\
                \hat{T}_{ij2}^{\alpha_{ij}} \ket{\{n\}} = -J \;\hat{b}_j^{\dagger}\hat{b}_i \;\delta(n_i-n_j+\alpha_{ij}-1) \ket{\{n\}} 
            \end{align}
        \end{subequations}

        Where Eqs.~\eqref{app:eq:Tij12} describe the inward ($\ell=1$) and outward ($\ell=2$) hopping respectively on the link comprising of sites $i$ and $j$. The change of the diagonal part of the energy due to hopping on the link connecting sites $i$ and $j$, under the action of the inward and the outward hopping operators are given by,
        
        \begin{subequations}
        \begin{align}
            \Delta\epsilon_{ij1}^{\alpha_{ij}} = +\alpha_{ij}U_0 - \mu_i + \mu_j   \\
            \Delta\epsilon_{ij2}^{\alpha_{ij}} = -\alpha_{ij}U_0 + \mu_i - \mu_j  
        \end{align}
        \end{subequations}
        
        We use the above equations to identify the parameters $\alpha_{ij}$. The links for which Eq.~\eqref{app:eq:alpha0} is satisfied are low-energy processes, in the following sense
        
        \begin{equation}
            \alpha_{ij}\;U_0 + \mu_i - \mu_j \sim \mathcal{O}(J) \ll \mathcal{O}(U_0)
            \label{app:eq:alpha0}
        \end{equation}
        
        This method of identification will be helpful in the systematic construction of the perturbative canonical transformation, particularly in the case of a strongly repulsive Bose-Hubbard model with spatially inhomogeneous chemical potentials. The effective Hamiltonian up to the second-order in perturbation theory is then given by
        
        \begin{equation}
            \hat{H}_\text{eff} = e^{i\hat{S}}\hat{H}e^{-i\hat{S}} \simeq \hat{H}_0 + \hat{T}_0 + [i\hat{S},\hat{T}_0] + \frac{[i\hat{S},\hat{T}_1]}{2!}
            \label{app:eq:Effective_Hamiltonian}
        \end{equation}
        
        where $i\hat{S}$ is determined by the condition $[i\hat{S},\hat{H}_0]=-\hat{T}_1$, which combined with the fact that $[\hat{H}_0,{\hat{T}}_{rr'\ell}^{\alpha_{rr'}}] \ket{\{n\}} = \Delta\epsilon_{rr'\ell}^{\alpha_{rr'}} \hat{T}_{rr'\ell}^{\alpha_{rr'}} \ket{\{n\}}$ results in the following action of the generator of the perturbative canonical rotation, $i\hat{S}$ on the occupation number basis 
        
            \begin{equation}
            \begin{split}
                i\hat{S}\ket{\{n\}} =& -J\sum_{r=1}^L \sum_{r'\in\text{nn}^{+}(r)} \frac{\hat{b}_r^\dagger \hat{b}_{r'}}{\Delta\epsilon_{rr'1}^{\alpha_{rr'}}} \delta(n_r-n_r'-\alpha_{rr'}+1) \\ &
                + \frac{\hat{b}_{r'}^\dagger \hat{b}_r}{\Delta\epsilon_{rr'2}^{\alpha_{rr'}}} \delta(n_r-n_r'+\alpha_{rr'}-1) \ket{\{n\}}
                \label{app:eq:iS_defn}
            \end{split}
            \end{equation}    
     We also note that in the above expression, $\alpha_{rr'} \ne \alpha_{rr'}^0$ as the effective Hamiltonian constructed from this generator is devoid of hopping processes with energy penalty of $\mathcal{O}({J/U_0})$ and $\text{nn}^{+}(r)$ are the nearest neighbor of site $r$ towards its right.
    As we shall show in the following appendices, the form \eqref{app:eq:iS_defn} is well suited for on-the-fly computations of expectation values such as $\langle \tilde\Psi_{\text{GW}}(\boldsymbol{f})|\hat{H}|\tilde\Psi_{\text{GW}}(\boldsymbol{f})\rangle$ or $\langle \tilde\Psi_{\text{SW}}(\boldsymbol{f})|\hat{H}|\tilde\Psi_{\text{SW}}(\boldsymbol{f})\rangle$ without requiring to construct the states or the operators explicitly, allowing us to simulate moderate system sizes.

    \section{Evaluation and optimization of ground-state energy \label{app:sec:evaluate_energy_functional}}

    This section contains details related to evaluation and optimization of the variational energy $\langle \tilde\Psi_{\text{GW}}(\boldsymbol{f}) | \hat{H} | \tilde\Psi_{\text{GW}}(\boldsymbol{f}) \rangle$. We first consider the translationally symmetric case of the clean system. We start by recalling that for a clean Bose-Hubbard chain of length $L$, the particle-number-projected Gutzwiller variational state reads

    \begin{equation}
        \ket{\tilde\Psi_{\text{GW}}(\boldsymbol{f})} = \mathcal{N}_{N_b} \hat{\mathcal{P}}_{N_b} \left[\bigotimes_{r=1}^L \left(\sum_{n_r=0}^{N_{\text{max}}}f^{n_r}\ket{n_r}_r\right)\right]
        \label{app:main:eq:fixed_nb_gutz_ansatz-app}
    \end{equation}

    where $\hat{\mathcal{P}}_{N_b}$ is the projection operator onto a desired number of total particle-number ($N_b$) sector and $\mathcal{N}_b$ is a normalization constant which ensures $\langle \tilde\Psi_{\text{GW}}(\boldsymbol{f})|\tilde\Psi_{\text{GW}}(\boldsymbol{f})\rangle = 1$. The variational ansatz \eqref{app:main:eq:fixed_nb_gutz_ansatz-app} has built-in translational symmetry and belongs to the zero quasi-momentum sector of the Hilbert space. We recall that, in the expression above, $\ket{n_r}_r$ denotes a state with $n_r$ bosons at site $r$ and $\{f^{n_r}\}$ are the corresponding Gutzwiller weights which satisfy $\sum_{n_r=0}^{N_{\text{max}}} |f^{n_r}|^2=1$. We now note that the state $\ket{\tilde\Psi_{\text{GW}}(\boldsymbol{f})}$ can be expressed as  

    \begin{equation}
        \ket{\tilde\Psi_{\text{GW}}(\boldsymbol{f})} = \mathcal{N}_{N_b} \sideset{}{'}\sum_{\left\{n\right\}}F(\left\{n\right\}) \ket{\left\{n\right\}}
    \end{equation}

    Where $F_{\{n\}}=\left(\prod_{r=1}^L f^{n_r}\right)$ is the product of Gutzwiller weights associated with each occupation configuration $\{n\}$, and the primed summation symbol $\sideset{}{'}\sum$ represents a sum over all occupation-number configurations $\left\{n\right\}$ which satisfy $\sum_r n_r=N_b$. Since in a clean system the Gutzwiller weights $f^{n_r}$ are site independent, there is a lot of symmetry -- configurations $\{n\}$ and $\{n'\}$ which are related to each other by any permutation of the sites, all have the same Gutzwiller weight i.e. $F_{\{n\}}=F_{\{n'\}}$. A subgroup of this symmetry, which is also easy to implement in numerical simulations, is the translation symmetry which we now consider in order to reduce the computational complexity associated with evaluation of the variational energy $\langle \tilde\Psi_{\text{GW}}(\boldsymbol{f}) | \hat{\mathcal{H}} | \tilde\Psi_{\text{GW}}(\boldsymbol{f}) \rangle$. Denoting the lattice-translation operator by one site as $\hat{\mathcal{T}}$ and introducing the zero quasi-momentum state $\ket{\{n\},k=0} \equiv \left(\sum_{r=0}^{L-1} \hat{\mathcal{T}}^r \ket{\left\{n\right\}}\right)/\sqrt{\mathcal{N}_{\left\{n_r\right\}}}$ we can express $\ket{\tilde\Psi_{\text{GW}}(\boldsymbol{f})}$ as follows

    \begin{equation}
        \ket{\tilde\Psi_{\text{GW}}(\boldsymbol{f})} = \mathcal{N}_{N_b} \sideset{}{''}\sum_{\left\{n\right\}} F_{\{n\}} \sqrt{\mathcal{N}_{\{n\}}}\;\frac{\mathcal{R}_{\{n\}}}{L} \ket{\{n\},k=0}
        \label{app:main:eq:fixed_nb_gutz_ansatz-txsymm}
    \end{equation}  

    In the above equation $\mathcal{N}_{\{n\}}$ is the normalization constant of the zero quasi-momentum state $\ket{\{n\},k=0}$ and $\mathcal{R}_{\{n\}}$ is the periodicity of the state $\ket{\{n\}}$ under the action of $\hat{\mathcal{T}}$ (i.e. $\hat{\mathcal{T}}^{\mathcal{R}_{\{n\}}}\ket{\{n\}}=\ket{\{n\}}$). The double primed summation symbol $\sideset{}{''}\sum_{}$ denotes a sum over the set of distinct occupation-number configurations $\{n_r\}$ which (i) are not related to each other via lattice translations and (ii) satisfy $\sum_r n_r=N_b$. The factors $\sqrt{\mathcal{N}_{\{n\}}}\;\mathcal{R}_{\{n\}}/L$ compensate for the over-counting of terms which appear due to the introduction of the zero quasi-momentum state $\ket{\{n\},k=0}$. This is because the zero quasi-momentum state consists of occupation-number configurations with all possible translations, and not only distinct translations. This new form of the ansatz \eqref{app:main:eq:fixed_nb_gutz_ansatz-txsymm} has the advantage that the states $\ket{\{n\},k=0}$ automatically form an orthonormal basis of smaller dimensionality compared to the entire Hilbert space, and the matrix elements of a translationally invariant Hamiltonian such as \eqref{eq:BHmodel-1D} can be easily written in this basis. The remaining unknown piece is the normalization constant $\mathcal{N}_{N_b}$ which is fixed by requiring that

    \begin{equation}
        \langle \tilde\Psi_{\text{GW}}(\boldsymbol{f})|\tilde\Psi_{\text{GW}}(\boldsymbol{f})\rangle = \mathcal{N}_{N_b}^2 \sideset{}{''}\sum_{\{n\}} F_{\{n\}} \;\mathcal{R}_{\{n\}} = 1
    \end{equation}

    Using the form Eq.\eqref{app:main:eq:fixed_nb_gutz_ansatz-txsymm} of the ansatz Eq.~\eqref{app:main:eq:fixed_nb_gutz_ansatz-app}, and recalling that for a 1D lattice $\mathcal{N}_{\{n\}}=L^2/\mathcal{R}_{\{n\}}$, we find that the expectation values of $\hat{H}_0$ and $\hat{T}$ with respect to the state $\ket{\tilde\Psi_{\text{GW}}(\boldsymbol{f})}$ are given by

    \begin{widetext}
        \begin{eqnarray}
            \langle\tilde\Psi_{\text{GW}}(\boldsymbol{f})|\hat{H}_0|\tilde\Psi_{\text{GW}}(\boldsymbol{f})\rangle = \mathcal{N}_{N_b}^2 \sideset{}{''}\sum_{\left\{n\right\}} |F_{\{n\}}|^2 \;\mathcal{R}_{\{n\}}\sum_{r=1}^L\left(-\mu_0 n_r +\frac{U_0}{2}n_r(n_r-1)\right) \label{app:eq:txsymm-diagonal-term} \\
            \langle \tilde\Psi_{\text{GW}}(\boldsymbol{f})|\hat{T}|\tilde\Psi_{\text{GW}}(\boldsymbol{f})\rangle = -J\;\mathcal{N}_{N_b}^2 \sideset{}{''}\sum_{\left\{n\right\}} \left(\sum_{r=1}^L \sum_{e=-1}^{+1} \sqrt{n_{r+e}}\sqrt{n_r+1} \; F_{\{n\}} F_{\{n\}(r,e)}^\star \mathcal{R}_{\{n\}}\right) \label{app:eq:txsymm-offdiagonal-term}
        \end{eqnarray}
    \end{widetext}

    In Eqs.~\eqref{app:eq:txsymm-diagonal-term},\eqref{app:eq:txsymm-offdiagonal-term}, the notation $\left\{n\right\}(r,e)$ denotes a new occupation-number configuration which is obtained by hopping one particle from site $r+e$ to site $r$ in the Fock state $\left\{n\right\}$. Calculating the above sum in a numerical implementation is simple, as one does not need to search for the state index of configuration $\{n\}(r,e)$ which is required for example in ED-type calculations. The only information regrading the configuration $\{n\}(r,e)$ is the product of Gutzwiller weights $F_{\{n\}}$ and the periodicity of the configuration $\mathcal{R}_{\{n\}}$, which can be computed once the basis of the Hilbert space is generated with appropriate symmetries.    

    When the chemical potential profile is spatially inhomogeneous, the simplified expressions Eqs.~\eqref{app:eq:txsymm-diagonal-term},\eqref{app:eq:txsymm-offdiagonal-term} are no longer valid. Instead one needs to evaluate the following expressions

    \begin{widetext}
        \begin{eqnarray}
            \langle\tilde\Psi_{\text{GW}}(\boldsymbol{f})|\hat{H}_0|\tilde\Psi_{\text{GW}}(\boldsymbol{f})\rangle = \sum_{\{n\}} \sum_{r=1}^L\left(-\mu_rn_r+\frac{U_0}{2}n_r(n_r-1)\right) |F_{\{n\}}|^2 \label{app:eq:nosymm-diagonal-term} \\
            \langle \tilde\Psi_{\text{GW}}(\boldsymbol{f})|\hat{T}|\tilde\Psi_{\text{GW}}(\boldsymbol{f})\rangle = -J\sum_{\left\{n\right\}} \left(\sum_{r=1}^L \sum_{e=-1}^{+1} \sqrt{n_{r+e}}\sqrt{n_r+1} \; F_{\{n\}} F_{\{n\}(r,e)}^\star \right) \label{app:eq:nosymm-offdiagonal-term}
        \end{eqnarray}
    \end{widetext}

    The above procedure can be straightforwardly generalized for the canonically transformed Gutzwiller ansatz scenario. Recall that after canonical transformation, the effective Hamiltonian is given by Eq.~\eqref{app:eq:Effective_Hamiltonian} and in this case, one needs to minimize $\langle \tilde \Psi_{\text{GW}}(\boldsymbol{f})|\hat{H}_{\text{eff}}|\tilde \Psi_{\text{GW}}(\boldsymbol{f})\rangle$. As before, by introducing the translationally invariant zero quasi-momentum basis states $\ket{\{n\}(k=0)}$, and using Eq.~\eqref{app:eq:iS_defn} for the matrix elements of the generator $i\hat{S}$, we can derive expressions similar to Eqs.~\eqref{app:eq:txsymm-diagonal-term},\eqref{app:eq:txsymm-diagonal-term} for the translationally symmetric case and Eqs.~\eqref{app:eq:nosymm-diagonal-term},\eqref{app:eq:nosymm-offdiagonal-term} for the inhomogeneous case.
    
    In all cases above, the desired variational energies parametrized by the Gutzwiller coefficients $\boldsymbol{f}$ can be expressed as summations over a relevant Hilbert space basis. Due to the strong repulsive interactions, i.e. $J,\mu_r,\kappa \ll U_0$ (see main text), it is already energetically very expensive to have 3 bosons simultaneously on some site, and thus taking the cutoff $N_{\text{max}}=3$ is sufficient. Hence for optimizing the energy expressions we fix the maximum number of bosons $N_{\text{max}}$ allowed on each site to be $3$, and the variational parameters $\{f\}=(f^0,f^1,f^2,f^3)$, in the translationally invariant case, can be written in terms of angles $\alpha,\beta,\gamma$ as

    \begin{subequations}
    \begin{align}
        f^0 &= \cos(\alpha) \\
        f^1 &= \sin(\alpha) \cos(\beta) \\
        f^2 &= \sin(\alpha) \sin(\beta) \cos(\gamma) \\
        f^3 &= \sin(\alpha) \sin(\beta) \sin(\gamma)
    \end{align}
    \end{subequations}

    with $\alpha\in[0,\pi],\beta\in[0,\pi],\gamma\in[0,2\pi]$. This parametrization of the Gutzwiller parameters $\boldsymbol{f}$ ensures that $\sum_{q=0}^3 |f|_q^2 = 1$. Although such a parametrization is not strictly necessary, we find that using this additional parameterization benefits the optimization procedure as this reduces the search space from a 4 dimensional to a 3 dimensional one with known and compact domains. The search space, for translationally invariant cases is three dimensional and thus the energy expectation values can be optimized efficiently using various standard optimization algorithms which are readily available. We find that for our purposes, particularly, when considering inhomogeneous cases, where the number of variational parameters are larger (namely 3 angles $\alpha_r,\beta_r,\gamma_r$ for each site $r=1,2,...,L$), the diversity enhanced particle swarm optimization algorithm (DNPSO) \cite{WANG2013119,4982999} is best suited among the ones we tested.

    \section{Constructing the TDVP EOMs \label{app:sec:TDVP_EOM_construction}}

    In this section, we outline our general strategy to construct the TDVP equations of motion for the different variational states used in the main text. We first note that in all cases, we can express the normalized variational ansatz state $\ket{\Psi(\boldsymbol{x})}$ as

    \begin{equation}
        \ket{\Psi(\boldsymbol{x})} = \frac{1}{\mathcal{N}(\boldsymbol{x})} \ket{\Phi(\boldsymbol{x})}
    \end{equation}

    where $\mathcal{N}(\boldsymbol{x})=\sqrt{\langle \Phi(\boldsymbol{x})|\Phi(\boldsymbol{x})\rangle}$. For real parameters $\boldsymbol{x} \in \mathbb{R}^M$ the TDVP equations of motion are

    \begin{equation}
        \sum_{b=1}^M \mathcal{G}_{ab}(\boldsymbol{x}(t)) x_b(t) = \mathcal{F}_a(\boldsymbol{x}(t)), \;\; \forall a =1,2,...,M
    \end{equation}

    with
    
    \begin{subequations}
    \begin{align}
        \mathcal{G}_{ab}(\boldsymbol{x}) & = 2 \mathfrak{Im}\left[\langle \partial_a \Psi(\boldsymbol{x}) | \partial_b \Psi(\boldsymbol{x}) \rangle\right] \\
        \mathcal{F}_{a}(\boldsymbol{x}) & = -2 \mathfrak{Re}\left[\langle \partial_a \Psi(\boldsymbol{x}) |\hat{H}|\Psi(\boldsymbol{x}) \rangle \right]
    \end{align}
    \label{supp:eq:TDVP_EOMs}
    \end{subequations}

    The gradients $\partial_a \ket{\Psi(\boldsymbol{x})}$ are required to be computed in order to evaluate $\mathcal{G}_{ab}(\boldsymbol{x})$ and $\mathcal{F}_a(\boldsymbol{x})$ appearing in Eq.~\eqref{supp:eq:TDVP_EOMs}, at any given point $\boldsymbol{x}$ belonging to the variational manifold and can be expressed in terms of the gradients of $\ket{\Phi(\boldsymbol{x})}$ and $\mathcal{N}(\boldsymbol{x})$ as follows

    \begin{equation}
            \partial_a \ket{\Psi(\boldsymbol{x})} = \frac{1}{\mathcal{N}(\boldsymbol{x})} \partial_a \ket{\Phi(\boldsymbol{x})} - \frac{1}{\mathcal{N}^2(\boldsymbol{x})} \partial_a \mathcal{N}(\boldsymbol{x}) \ket{\Phi(\boldsymbol{x})}
            \label{app:eq:partial_a_Psi}
    \end{equation}
    
    As the components of $\ket{\Phi(\boldsymbol{x})}$ in the Fock basis are products of Gutzwiller coefficients $f_r^{n_r}$, the gradients can be expressed in terms of the powers of these Gutzwiller coefficients. 

    \subsection{Translationally symmetric particle-number projected Gutzwiller state}
    
    For a system with a homogeneous chemical potential profile, described variationally by a particle-number projected Gutzwiller ansatz we have  

    \begin{equation}
        \ket{\tilde \Phi(\boldsymbol{x})} = \sideset{}{'}\sum_{\{n\}} A_{\{n\}}(\boldsymbol{x}) \ket{\{n\}}
        \label{app:eq:TDVP_Phi_txsymm}
    \end{equation}
    
    With $M=2(N_{\text{max}}+1)$ being the total number of real variational parameters. The Gutzwiller parameters $\{f^{n_r}\}$, are now required to be complex valued for their use in quantum dynamics, and they are related to the $M$ real parameters $x_1,...,x_M$ through the following identification

    \begin{subequations}
        \begin{align}
            (x_1,x_2,...,x_{M/2}) & \equiv (\mathfrak{Re}(f^0),...,\mathfrak{Re}(f^{N_{\text{max}}})) \nonumber \\
            (x_{M/2+1}x_{M/2+2},...,x_{M}) & \equiv (\mathfrak{Im}(f^0),...,\mathfrak{Im}(f^{N_{\text{max}}})) \nonumber
        \end{align}
    \end{subequations}
        
    In Eq.~\eqref{app:eq:TDVP_Phi_txsymm}, the coefficients $A_{\{n\}}(\boldsymbol{x})$ take the following form

    \begin{equation}
        A_{\{n\}}(\boldsymbol{x}) = \prod_{n_b=0}^{N_{\text{max}}} \left(f^{n_b}\right)^{c_{n_b,\{n\}}}
        \label{app:eq:partial_A_homogeneous}
    \end{equation}

    where $c_{n_b,\{n\}} \in \{0,1,...L\}$ are integers which count the number of sites which have $n_b$ bosons in the Fock state $\ket{\{n\}}$. The gradients $\partial_a A_{\{n\}}(\boldsymbol{x})$ are given by

    \begin{equation}
        \partial_a A_{\{n\}}(\boldsymbol{x}) = \xi_a c_{a,\{n\}}\left(f^{a}\right)^{c_{a,\{n\}}-1}(1-\delta_{c_{a,{\{n\}}},0})  \prod_{n_b\ne a} \left(f^{n_b}\right)^{c_{n_b,\{n\}}}
        \label{app:eq:grad_A}
    \end{equation}

    with $\xi_a=1$ for $a<M/2$ and $\xi_a=i$ otherwise. Using the above expression, we can find the gradients $\partial_a \ket{\Phi(\boldsymbol{x})}$ and also $\partial_a \mathcal{N}(\boldsymbol{x})$ using the relation 

    \begin{equation}
        \partial_a \mathcal{N} = \mathfrak{Re}\left( \langle \partial_a\Phi|\Phi \rangle \right)/\mathcal{N}(\boldsymbol{x})
    \end{equation}

    Now we are in a position to compute $\partial_a\ket{\Psi(\boldsymbol{x})}$ via Eq.~\eqref{app:eq:partial_a_Psi} and thus $\mathcal{G}_{ab}(\boldsymbol{x})$ and $\mathcal{F}_a(\boldsymbol{x})$, which allows us to evolve the TDVP equations of motion for the real parameters $\boldsymbol{x}(t)$ using a numerical algorithm. \newline

    \subsection{Inhomogeneous particle-number projected Gutzwiller state}
    
    We now briefly describe how the aforementioned procedure is adapted for inhomogeneous cases. In this case one needs $2L(N_{\text{max}}+1)$ number of real variational parameters with the identification

    \begin{subequations}
        \begin{align}
            (x_1,x_2,...,x_{M/2}) & \equiv (\mathfrak{Re}(f_1^0),...,\mathfrak{Re}(f_L^{N_{\text{max}}})) \nonumber \\
            (x_{M/2+1}x_{M/2+2},...,x_{M}) & \equiv (\mathfrak{Im}(f_1^0),...,\mathfrak{Im}(f_L^{N_{\text{max}}})) \nonumber
        \end{align}
    \end{subequations}

    We now proceed as before by noting that in this case the state $\ket{\tilde\Phi(\boldsymbol{x})}$ in the computational basis reads

    \begin{equation}
        \ket{\tilde\Phi(\boldsymbol{x})} = \sideset{}{'}\sum_{\{n\}} A_{\{n\}}(\boldsymbol{x}) \ket{\{n\}} \\
    \end{equation}

    where the $A_{\{n\}}(\boldsymbol{x})$'s now are

    \begin{equation}
        A_{\{n\}}(\boldsymbol{x}) = \prod_{r=1}^L \left(f_r^{n_r}\right)^{c_{r,n_r,\{n\}}}
        \label{supp:eq:A_general}
    \end{equation}

    Here $c_{r,n_r,\{n\}}$'s are integers which take the value 1, if the site $r$ in the Fock state $\ket{\{n\}}$ have $n_r$ bosons and is 0 otherwise. Henceforth we set the convention that $n_r$ stands for the number of bosons at site $r$ of the Fock state $\ket{\{n\}}$. 
    From Eq.~\eqref{supp:eq:A_general}, we find that the gradients $\partial_a A_{\{n\}}(\boldsymbol{x})$ are expressed as
    \begin{equation}
        \partial_a A_{\{n\}}(\boldsymbol{x}) = \xi_a \prod_{r\ne(r)_a} (1-\delta_{c_{r,n_r,\{n\}},0}) (f_r^{n_r})^{c_{r,n_r,\{n\}}}
        \label{app:eq:partial_A_inhomogeneous}
    \end{equation}

    The factor $\xi_a$ has the same meaning as in Eq.~\eqref{app:eq:grad_A} and $(r)_a$ stands for the site index associated with the $a^\text{th}$ variational parameter. With the help of Eq.~\eqref{app:eq:partial_A_inhomogeneous} we can again construct $\mathcal{G}_{ab}(\boldsymbol{x})$, $\mathcal{F}_a(\boldsymbol{x})$ for any $\boldsymbol{x}$ and solve the resulting TDVP equations of motion. 

    \subsection{Translationally symmetric particle-number projected Gutzwiller state with additional correlations induced via $e^{i\hat{S}}$}
    
    In the analysis of the quantum Mpemba effect in the main text, we recall that the variational state reads (upto normalization)

    \begin{equation}
        \ket{\Psi(g,\boldsymbol{f})} \sim (1+ig\hat{S}) \ket{\tilde \Psi_{\text{GW}}(\boldsymbol{f})}
    \end{equation}
    
    For this ansatz the number of variational parameters are $M+1$, with $M=2(N_{\text{max}}+1)$. In this case we again start with the identification that
    
    \begin{subequations}
        \begin{align}
            (x_1,x_2,...,x_{M/2}) & \equiv (\text{Re}(f^0),...,\text{Re}(f^{N_{\text{max}}}))  \nonumber \\
            (x_{M/2+1}x_{M/2+2},...,x_{M}) & \equiv (\mathfrak{Im}(f^0),...,\mathfrak{Im}(f^{N_{\text{max}}})) \nonumber \\
            x_{M+1} & \equiv g
        \end{align}
    \end{subequations}    
    
        Using the notation used throughout this section before, we find

        \begin{equation}
            \ket{\Phi(\boldsymbol{x})} = \sideset{}{'}\sum_{\{n\}} \left( A_{\{n\}}(\boldsymbol{x}) + x_{M+1} B_{\{n\}}(\boldsymbol{x}) \right) \ket{\{n\}}
        \end{equation}

        where $A_{\{n\}}(\boldsymbol{x})$'s are now the same as in Eq.~\eqref{app:eq:partial_A_homogeneous}. The new coefficients $B_{\{n\}}(\boldsymbol{x})$'s can be determined by the action of $i\hat{S}$ on $\ket{\Phi(\boldsymbol{x})}$. For the homogeneous case, and with the matrix elements of $i\hat{S}$ in the Fock basis given by Eq.~\eqref{app:eq:iS_defn}, we find that 

        \begin{widetext}
        \begin{equation}
            B_{\{n\}}(\boldsymbol{x}) = -\frac{J}{U_0}\sum_{r=1}^L \left(\prod_{r' \ne r,r+1} f_{n_{r'}}\right) \frac{\sqrt{n_{r+1-}\left(n_r+1\right)}}{n_{r+1}-1-n_r} f_{n_r+1} f_{n_{r+1}-1} + \frac{\sqrt{n_{r}\left(n_{r+1}+1\right)}}{n_{r}-1-n_{r+1}} f_{n_r-1} f_{n_{r+1}+1}
        \end{equation}    
        \end{widetext}
        
        Now the gradients $\partial_a \ket{\Phi(\boldsymbol{x})}$, $\forall a=1,M$ are,  

        \begin{equation}
            \partial_a \ket{\Phi(\boldsymbol{x})} = \sideset{}{'}\sum_{\{n\}} \left( \partial_a A_{\{n\}}(\boldsymbol{x}) + x_{M+1}  \partial_a B_{\{n\}}(\boldsymbol{x}) \right) \ket{\{n\}}
        \end{equation}

       while for $a=M+1$,

        \begin{equation}
            \partial_a \ket{\Phi(\boldsymbol{x})} = \sideset{}{'}\sum_{\{n\}} B_{\{n\}}(\boldsymbol{x}) \ket{\{n\}}
        \end{equation}

        Using the above equation we can now compute the TDVP equations of motion and solve them numerically.\newline
        
        In all the cases considered in this section, once we have solved the TDVP equation of motion starting with some desired initial condition, we have found the trajectory $\boldsymbol{x}(t)$ in the $M$ dimensional real parameter space. From this solution we can find the instantaneous quantum states, compute the time-evolution of desired observables and compare those with exact evolution as done in the main text. 

\end{appendix}

\bibliography{references}

\end{document}